\documentclass[twocolumn]{aastex62}

\usepackage{longtable}
\usepackage{soul}

\usepackage{natbib}
\usepackage{graphicx}
\newcolumntype{H}{>{\setbox0=\hbox\bgroup}c<{\egroup}@{}}

\graphicspath{{./}{figures/}}

\accepted{to ApJS on February 17, 2021}

\shorttitle{101 Trojans with K2}
\shortauthors{Kalup et al.}
\begin{document}

\title{101 Trojans: a tale of period bimodality, binaries, and extremely slow rotators from K2 photometry}

\correspondingauthor{Csilla E. Kalup}
\email{kalup.csilla@csfk.org}

\author{Csilla E. Kalup}
\affiliation{Konkoly Observatory, Research Centre for Astronomy and Earth Sciences, Konkoly Thege 15-17, H-1121 Budapest, Hungary}
\affiliation{Department of Astronomy, E\"otv\"os Lor\'and University, P\'azm\'any P\'eter s\'et\'any 1/A, H-1171 Budapest, Hungary}
\affiliation{MTA CSFK Lend\"ulet Near-Field Cosmology Research Group}

\author[0000-0002-8159-1599]{L\'aszl\'o Moln\'ar}
\affiliation{Konkoly Observatory, Research Centre for Astronomy and Earth Sciences, Konkoly Thege 15-17, H-1121 Budapest, Hungary}
\affiliation{ELTE E\"otv\"os Lor\'and University, Institute of Physics, 1117, P\'azm\'any P\'eter s\'et\'any 1/A, Budapest, Hungary}
\affiliation{MTA CSFK Lend\"ulet Near-Field Cosmology Research Group}

\author[0000-0002-8722-6875]{Csaba Kiss}
\affiliation{Konkoly Observatory, Research Centre for Astronomy and Earth Sciences, Konkoly Thege 15-17, H-1121 Budapest, Hungary}
\affiliation{ELTE E\"otv\"os Lor\'and University, Institute of Physics, 1117, P\'azm\'any P\'eter s\'et\'any 1/A, Budapest, Hungary}

\author[0000-0002-0606-7930]{Gyula M. Szab\'o}
\affiliation{ELTE E\"otv\"os Lor\'and University, Gothard Astrophysical Observatory, 9700 Szombathely, Szent Imre h. u. 112, Hungary}
\affiliation{MTA-ELTE Exoplanet Research Group, 9700 Szombathely, Szent Imre h. u. 112, Hungary}

\author[0000-0001-5449-2467]{Andr\'as P\'al}
\affiliation{Konkoly Observatory, Research Centre for Astronomy and Earth Sciences, Konkoly Thege 15-17, H-1121 Budapest, Hungary}
\affiliation{Department of Astronomy, E\"otv\"os Lor\'and University, P\'azm\'any P\'eter s\'et\'any 1/A, H-1171 Budapest, Hungary}

\author[0000-0002-1698-605X]{R\'obert Szak\'ats}
\affiliation{Konkoly Observatory, Research Centre for Astronomy and Earth Sciences, Konkoly Thege 15-17, H-1121 Budapest, Hungary}

\author[0000-0003-0926-3950]{Kriszti\'an S\'arneczky}
\affiliation{Konkoly Observatory, Research Centre for Astronomy and Earth Sciences, Konkoly Thege 15-17, H-1121 Budapest, Hungary}

\author[0000-0001-8764-7832]{J\'ozsef Vink\'o}
\affiliation{Konkoly Observatory, Research Centre for Astronomy and Earth Sciences, Konkoly Thege 15-17, H-1121 Budapest, Hungary}

\author[0000-0002-3258-1909]{R\'obert Szab\'o}
\affiliation{Konkoly Observatory, Research Centre for Astronomy and Earth Sciences, Konkoly Thege 15-17, H-1121 Budapest, Hungary}
\affiliation{MTA CSFK Lend\"ulet Near-Field Cosmology Research Group}
\affiliation{ELTE E\"otv\"os Lor\'and University, Institute of Physics, 1117, P\'azm\'any P\'eter s\'et\'any 1/A, Budapest, Hungary}

\author[0000-0002-9511-0901]{Vikt\'oria Kecskem\'ethy}
\affiliation{Konkoly Observatory, Research Centre for Astronomy and Earth Sciences, Konkoly Thege 15-17, H-1121 Budapest, Hungary}
\affiliation{Department of Astronomy, E\"otv\"os Lor\'and University, P\'azm\'any P\'eter s\'et\'any 1/A, H-1171 Budapest, Hungary}

\author[0000-0002-3234-1374]{L\'aszl\'o L. Kiss}
\affiliation{Konkoly Observatory, Research Centre for Astronomy and Earth Sciences, Konkoly Thege 15-17, H-1121 Budapest, Hungary}
\affiliation{MTA CSFK Lend\"ulet Near-Field Cosmology Research Group}

\begin{abstract}
Various properties of Jovian trojan asteroids such as composition, rotation periods, and photometric amplitudes, or the rate of binarity in the population can provide information and constraints on the evolution of the group and of the Solar System itself. 
Here we present new photometric properties of 45 Jovian trojans from the K2 mission of the \textit{Kepler} space telescope, and present phase-folded light curves for 44 targets, including (11351) Leucus, one of the targets of the \textit{Lucy} mission. 
We extend our sample to 101 asteroids with previous K2 Trojan measurements, then compare their combined amplitude- and frequency distributions to other ground-based and space data. We show that there is a dichotomy in the periods of Trojans with a separation at $\sim 100$ hr. 
We find that 25\% of the sample are slow rotators (P$\geq$30 hr), which excess can be attributed to binary objects. We also show that 32 systems can be classified as potential detached binary systems. 
Finally, we calculate density and rotation constraints for the asteroids. Both the spin barrier and fits to strengthless ellipsoid models indicate low densities and thus compositions similar to cometary and TNO populations throughout the sample. This supports the scenario of outer Solar System origin for Jovian trojans.

\end{abstract}


\section{Introduction} \label{sec:intro}
Jovian trojan asteroids are located around the L4 and L5 Lagrange points of the Sun-Jupiter system, in 1:1 mean-motion resonance. According to the Minor Planet Center\footnote{\url{https://minorplanetcenter.net/iau/lists/JupiterTrojans.html}}, at the time of writing we know 8190 Jovian trojans, but of those in the Asteroid Light Curve Database\footnote{\url{http://www.minorplanet.info/lightcurvedatabase.html}}, only $\sim$ 5\% have rotational period and/or amplitude information \citep[LCDB,][]{LCDB}. However, it is possible that there are more than half a million asteroids out there with diameter of $>$1 km, which is comparable to the number of main belt asteroids \citep{Yoshida2005}. The number of known asteroids is complete down to $H<13$\,mag absolute brightness (i.e. having $d \gtrsim 15$ km diameter) \citep{Vinogradova2015}, and based on SDSS measurements, asteroids larger than 10 km are all identified \citep{Szabo2007}. 

The surface characteristics of Jovian trojans show similarities to those of trans-Neptunian objects \citep{Fraser2014,Emery2015,Wong2016}, and dynamical models also indicate that they had likely formed in the outer planetesimal disk, then were implanted to their current orbits after planetary encounters \citep{Nesvorny2013,Pirani2019}. Rotational characteristics, obtained via time-series photometry, also indicate that the Trojans share more similarities with objects in the outer Solar System than with main belt asteroids \citep[see,e.g.,][]{French2015}.

Recently, time-resolved photometry of asteroids have become possible with space-based instruments, too: both the \textit{Kepler} and TESS missions have encountered numerous objects within the Solar System while observing stellar targets. TESS can observe mostly objects within the main belt, but the larger aperture of {\it Kepler} allowed the measurement of fainter and farther objects, such as Hildas, Trojans, centaurs and TNOs during its K2 mission. The K2 mission consisted of 3 months long Campaigns around the ecliptic plane, which provided a unique opportunity to observe asteroids from space \citep{Howell2014}. These high quality, several-week-long, uninterrupted and continuous measurements are free from aliases, thus, very suitable for studying asteroids with rotational periods longer than 10 or even 100 hours.
In contrast, rotational statistics from previous ground-based measurements could be strongly biased, favouring the detection of short (P\,$\lesssim$\,12\,hr) rotation periods. More recently, a large number of slow rotators were identified in various small body populations, including Centaurs \citep{Marton2020}, Jovian trojans \citep{ryan2017,Szabo2017} and Hildas \citep{Szabo2020ApJS} using K2 data, and also among main belt asteroids, based on TESS measurements \citep{Pal2020ApJS}.

One of the possible explanations of an excess of slow rotators can be binarity. The importance of determining the binary fraction of asteroids in a certain population is that it can provide constraints on the dynamical evolution of the Solar System. Using mid-infrared WISE data, \cite{Sonnett2015} derived that 13\%--23\% of Trojans are extremely elongated or binaries. Based on the \citet{ryan2017} and \cite{Szabo2017} Jovian trojan sample \citet{nesvorny2020} suggested that the large fraction of $\sim$15\% of very slow rotators in that sample with periods over 100\,hr can be explained by tidally synchronized binaries, originally formed as equal-sized binaries in the massive outer disk at 20-30\,au in the early Solar system.

\cite{Szabo2017} analyzed 56 Jovian trojans from Campaign 6 (C6). In this work, we expand that K2 Trojan sample to 101 asteroids. There are two notable targets in this paper. The first one is (11351) Leucus, which belongs to the extremely slow rotators, and it was chosen as one of the flyby targets of the \textit{Lucy} mission \citep{Levison2016}. The spacecraft is planned to visit the asteroid in April 2028. The second notable target is (13062) Podarkes, which is the principal body of the proposed Podarkes family that belongs to the larger Menelaus clan \citep{Roig2008}. 

In the framework of the K2 mission our group already published results concerning various objects, from main belt asteroids \citep{Szabo2015, Szabo2016, Molnar2018} to trans-Neptunian objects \citep{Pal2015, Pal2016}, Centaurs \citep{Marton2020} and irregular moons \citep{Kiss2016, FT2017}. K2 light curves were also used to model main belt asteroids by \citet{marciniak2019} and \citet{Podlevska2020}. In this work we continue to explore the K2 data from the Solar System. In Sections \ref{sect:obs} and \ref{sect:data} we describe the observations made by \textit{Kepler} and the data reduction steps. In Section \ref{sect:res} and \ref{sect:analysis} we present the results obtained from the photometry and the analysis of the rotational periods and light curve amplitudes of the asteroids. Finally, a summary is provided in Sect.~\ref{sect:summary}.

\begin{figure*}
\includegraphics[width=0.98\textwidth]{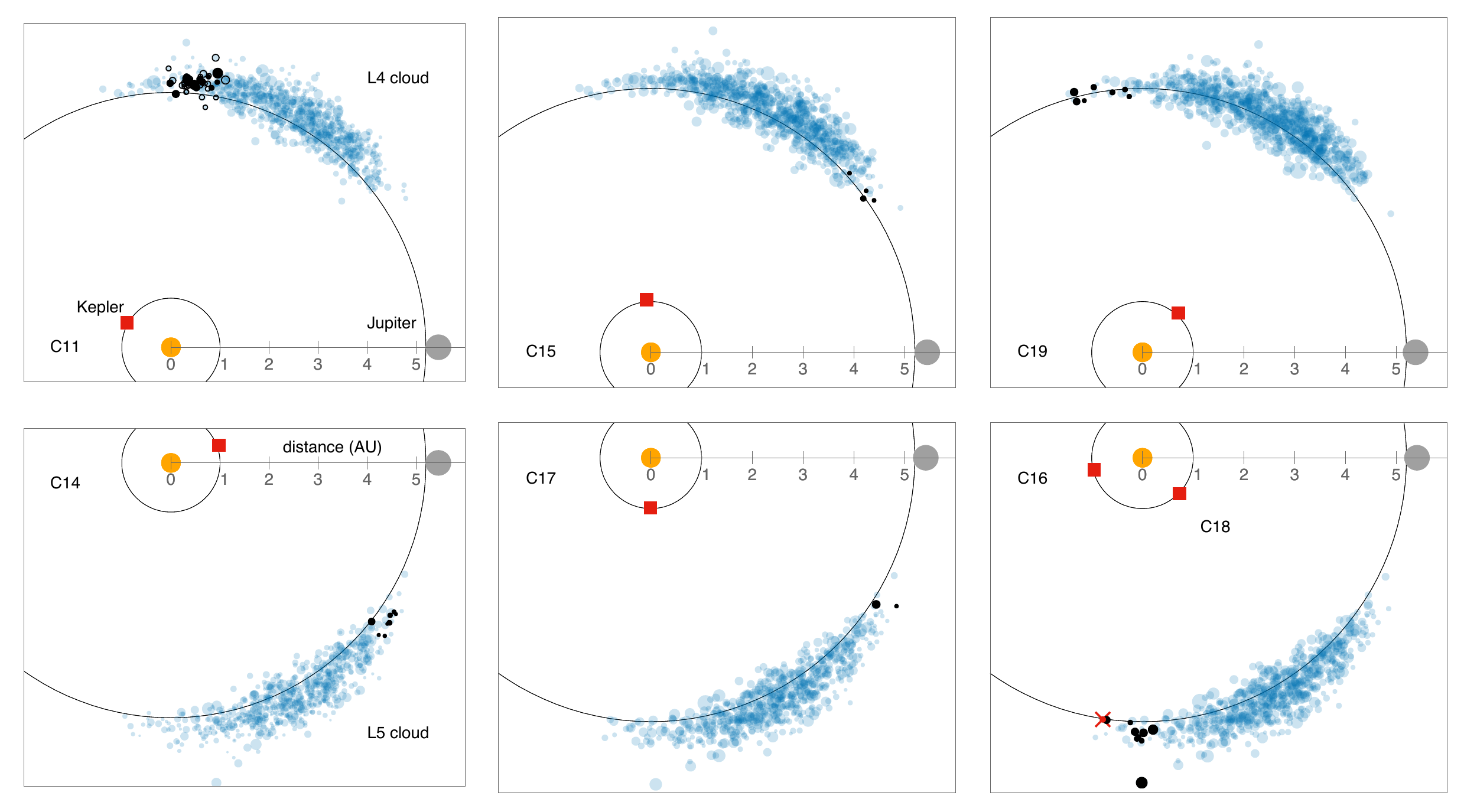}
\caption{Positions of the Trojan swarms and the observed targets in the L4 (top row) and L5 (bottom row) clouds in the Solar System relative to the Sun and Jupiter, in each Campaign. The position of \textit{Kepler} is also marked. Sizes of dots indicate apparent brightness as seem from \textit{Kepler}. Black dots indicate observed targets; black circles in C11 are targets that were observed but were too faint to produce usable light curves; the object marked with red cross in C16/C18 is 2001 SC101, observed in both Campaigns. Data obtained from the JPL Horizons service.}
\label{fig:maps}
\end{figure*}

\section{Observations}
\label{sect:obs}
In this paper, we present light curves and photometric properties of 45 Jovian trojan asteroids, observed by \textit{Kepler} during K2 Campaigns 11-19, between 2016 October 21.27 and 2018 September 23.10, in long cadence mode (29.41 min sampling frequency). The apparent trajectories allocated for these targets were observed via \textit{K2} Guest Observer proposals. The target pixel time series files have been retrieved from the Mikulsi Archive for Space Telescopes\footnote{\url{http://archive.stsci.edu/}} (MAST). The main parameters of the 45 observed Jovian trojans, including the duty cycles for each object, are summarized in Table \ref{tab:obs} in the Appendix. Unlike C6 that looked into the dense regions of the L4 group, later campaigns captured only the edges of the L4 and L5 clouds, as shown in Fig.~\ref{fig:maps}. Therefore, especially after C11, only a handful of objects were observed in each run.

Two Campaigns required special treatment. C11 was aimed at the Galactic bulge area, thus, the Trojans were followed in front of a very dense stellar background. Therefore, even though 38 targets were observed in C11, many of them were too faint to be detectable against the stellar background, hence we were not able to collect meaningful photometry. Due to this circumstance, we prioritized targets that were either bright with respect to the background stars or moved in front of less dense stellar fields.  Our results from C11 generally agree with the light curves produced by the K2 Moving Bodies Project\footnote{\url{https://christinahedges.github.io/asteriks/index.html}}, which, at the time of writing this paper, included Solar System data processed up to C13.

The other Campaign that needed special attention was the last one, C19. \textit{Kepler} operated right until the spacecraft  ran out of fuel. Due to the drop in fuel pressure, C19 suffered from degraded pointing accuracy, then ended prematurely. Only a roughly two-week long segment contains useful data that was taken with nominal pointing. Serendipitously, these two week coincided with the time when the Trojans selected for observation in C19 traversed the pixels placed along their tracks, making it possible to measure their brightness variations precisely.

The data we use in this paper were proposed through K2 GO programs GO11051, GO14085, GO15085, GO17028, GO18028 and GO19028\footnote{\url{https://keplerscience.arc.nasa.gov/k2-approved-programs.html}}.

\section{Data reduction}
\label{sect:data}
The data reduction steps of K2 long cadence data for moving objects have been analogous to other earlier analysis of asteroid targets that already have been discussed by our group in previous works \citep[see, e.g.,][]{Szabo2017,Molnar2018,Marton2020}. 
To process Kepler observations, we have utilized the FITSH software package and our external scripts. In brief, we assembled mosaic images from the individual Target Pixel Files of the target and some nearby stars, and derived the astrometric solution to register them into the same reference system. Then, a master image was created 
from a few dozens of frames that did not contain the target, which was then subtracted from all the frames. After that step only signals belonging to moving and variable features remained on each frame.

We generated the coordinates of asteroids from JPL Horizons ephemerides\footnote{\url{https://ssd.jpl.nasa.gov/?horizons}}. We converted fluxes to magnitude based on USNO \textit{R} magnitudes of the nearby stars \citep{Monet2003}. We note that we compute the same scaling law between \textit{Kepler} fluxes and \textit{R} magnitudes that was determined for \textit{Kp} magnitudes by \citet{Lund2015}. 

\subsection{Campaigns 11 and 19}
C11 was split into two parts due to an initial pointing error. Many of the Trojan targets were observed during the first half (C111) for which we did not have an astrometric solution readily available. For those we used the nominal coordinates of the neighboring TPFs and generated the astrometric matches manually. We then enlarged the mosaic images of all C11 targets by $\sim 3$ times, distributing the flux levels from the original pixels among the new ones. This method can help to reduce the fringing on the residual images which is caused by the sub-pixel position differences between the master image and the registered individual frames. Moreover, we are able to use tighter apertures just as we did for the Centaur light curves in \citet{Marton2020}. These tight apertures did not cover the edges of the PSFs, which led to slight flux loss in some cases. We corrected the light curves when necessary by matching them to the average brightness and amplitude measured on the non-enlarged images. The differences in average brightness were between 0.05--0.3 mag for most targets. In three cases we restricted the aperture to the center of the PSF which required corrections between 0.5--0.75 mag. Overall, we were able to derive rotation periods for 14 objects out of 38 observed in this campaign.

For C19 we generated the astrometry manually again, based on the nominal TPF coordinates. We tested enlarged pixels for this C19 too but the method did not offer significant improvements. We decided to use non-enlarged pixels in this campaign.

\subsection{Post--processing}
Obvious outliers were filtered out from the light curves after finishing the photometry. We then compared each light curve to the differential images and also filtered out data points that were affected by excess noise from stellar residuals, CCD crosstalk patterns or partial coverage of the PSF. We calculated the duty cycles for each target to quantify how many data points were used in our analysis from each observing window. Values, along with Campaign numbers, length, start and end times are listed in Table \ref{tab:obs} in the Appendix. A sample of the data file containing measurements of all asteroids is shown in Table~\ref{tab:lc}.

Finally, we created rectified versions of each light curve for the period search. We subtracted a linear fit from all light curves except for three targets in order to take into account the slow changes in brightness caused by the changing phase angle and distances to the targets. In two cases we applied no corrections, and in one case a quadratic fit was required.

\begin{deluxetable*}{lcccccccccc}
\tablecaption{Sample table of the photometry of Trojan asteroids observed in Campaigns 11 to 19 K2 mission. The table includes the identification numbers of the asteroids, the measured \textit{Kp} brightness and uncertainty from each frame, plus the corresponding equatorial coordinates (RA and Dec), heliocentric ecliptic longitudes and latitudes ($\lambda$ and $\beta$), observer--target range ($\Delta$), solar elongation (SOT), and phase angle (STO) values. The entire table is available online.
\label{tab:lc}}
\tablehead{
\colhead{ID} & \colhead{JD\_UTC [day]} & \colhead{Kp [mag]}  & \colhead{$\sigma_{Kp}$ [mag]} & \colhead{RA [deg]} & \colhead{DEC [deg]} & \colhead{$\lambda$ [deg]} & \colhead{$\beta$ [deg]} & \colhead{$\Delta$ [AU]} & \colhead{SOT [deg]} & \colhead{STO [deg]}}
\startdata
1871	&	2458268.868974	&	18.9660	&	0.0131	&	132.99263	&	16.51030	&	139.94735	&	-0.87773	&	4.801031	&	122.60730	&	9.10594	\\ 
1871	&	2458268.889407	&	18.8972	&	0.0127	&	132.99208	&	16.51075	&	139.94890	&	-0.87750	&	4.801292	&	122.58714	&	9.10812	\\ 
1871	&	2458268.909841	&	18.9003	&	0.0113	&	132.99154	&	16.51120	&	139.95045	&	-0.87727	&	4.801554	&	122.56699	&	9.11030	\\ 
1871	&	2458268.930275	&	18.8511	&	0.0133	&	132.99099	&	16.51166	&	139.95201	&	-0.87703	&	4.801816	&	122.54684	&	9.11248	\\ 
1871	&	2458268.950708	&	18.8886	&	0.0150	&	132.99045	&	16.51211	&	139.95356	&	-0.87680	&	4.802077	&	122.52669	&	9.11466	\\ 
\multicolumn{11}{l}{\dots}\\
\enddata
\end{deluxetable*}

\subsection{Period search}

The final light curves were analyzed with two different methods. First, we used our own residual minimization algorithm \citep[see][for previous studies]{Pal2016,Molnar2018,Marton2020}. In this case, we fit the data $m(\Delta t)$ with a function
\begin{equation}
m(\Delta t) ~=~ A + B \cos(2\pi f\Delta t) + C \sin(2\pi f\Delta t)
 \label{eq:freq}
\end{equation}
where $f$ is a trial frequency, and $\Delta t = T - t$, where $T$ is the approximate center of the time series. 

The dispersion of the light curve residual as a function of the trial frequency, $S(f)$, is computed by scanning the interval of the physically meaningful frequencies with a step size $\Delta f=0.002$ d$^{-1}$ to determine the parameters $A$, $B$ and $C$ via Eq.~\ref{eq:freq}. Then we searched for the minimum of the $S(f)$ dispersion curves, and the frequency corresponding to the minimum was considered as the best-fit one ($f_{b}$). Note that the best-fit frequencies obtained with this method give the same result as the Lomb-Scargle periodogram or Fast Fourier Transform methods \citep[see][]{Molnar2018}. After the automatic processing of the period search, we phase-folded the light curves with the best period, $P_b = f_b^{-1}$ and $2 P_b$, and inspected all the phased light curves visually. It turned out that folding with the double period gave the better fit for every Trojan in this paper. We also binned the phased light curves to see the shape of light curves better, and obtained the full amplitudes as the difference between the maximum and minimum of the binned light curve. The main frequencies were also determined with the FAMIAS code \citep{Zima2008}. The resulting frequencies from FAMIAS were identical to our ones obtained above. The frequency uncertainties were calculated with FAMIAS.

\section{Results}
\label{sect:res}

\begin{figure*}
\includegraphics[width=0.98\textwidth]{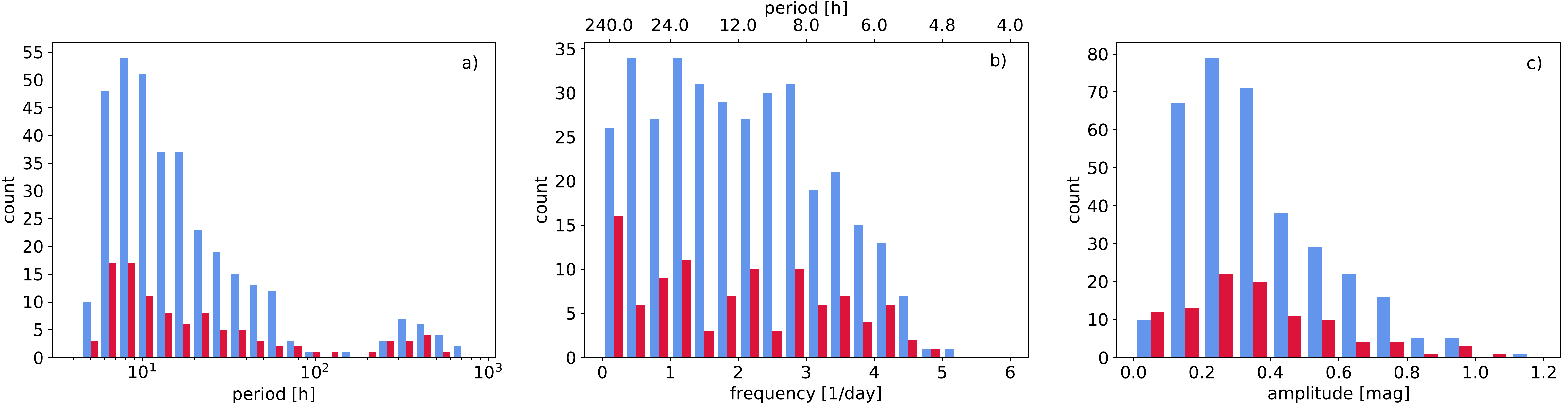}
\caption{(a) and (b): Period and frequency distributions of K2 Jovian trojan asteroids (red bars), compared to the period and frequency distributions of Jovian trojans from LCDB (blue bars). (c): K2 and LCDB amplitude distributions, using the same color coding. Where the LCDB also included a minimum amplitude we used the average of the maximum and minimum observed amplitudes.}
\label{fig:hists}
\end{figure*}

\subsection{Rotation periods and amplitudes}
We present the photometric properties of our sample in Table \ref{tab:data}. The period and amplitude values span a wide range. The shortest period is 5.124 hr ((151883) 2003 WQ25), while the longest one is 445.73 hr ((11351) Leucus). The smallest amplitude is 0.065 mag ((116567) 2004 BV84), and the largest one is 0.9644 mag ((76820) 2007 RW105). The median values of the periods and amplitudes are 8.931 hr and 0.3087 mag, respectively.

We were able to determine rotation periods and amplitudes for all but one target. For (60421) 2000 CZ31 we did not detect any significant coherent frequency components in the frequency spectrum. We determined the average upper limit for the rotation amplitude as $\lesssim 0.05$~mag at an expected period of 12 hours. Since low-frequency noise is present in the spectrum, this upper limit increases to 0.08~mag for periods longer than one day.

We present the period, frequency and amplitude distributions of the overall sample of Jovian trojans from all K2 Campaigns (i.e., our sample complemented with Trojans from C6). In \cite{Szabo2017}, asteroid 65227 has two periods in the residual spectrum, and we used their double peak solutions (7.06 hr and 49.7 hr) in the later analysis. We note that the presence of two periods in the spectrum can be a signal of possible binarity, therefore we consider these periods as the rotational periods of the two components: the longer period is the orbit of the secondary (which is often tidally locked) and the shorter period is the rotation of the primary (when not tidally locked). We also note that in the case of 23958, there was a factor--of--2 error in Table A.2, the good period value is the half of the presented value (571.425 hr).

Panel (a) and (b) of Fig.~\ref{fig:hists} show the period and frequency distributions, respectively. Red bars represent the overall K2 Jovian trojan sample, blue bars represent the Jovian trojan asteroids obtained from the LCDB. Fig.~\ref{fig:hists}a shows that 38\% of the K2 Trojan periods are shorter than 10 hr, while 12\% of the sample has very long rotational periods, between 100--600 hr. The median of the overall sample (12.65 hr) is the same as the median of Trojans from LCDB (12.75 hr). In panel (b) of Fig.~\ref{fig:hists} we present the frequency distributions of Jovian trojans. It shows a similar cut off at short periods as the LCDB sample. In Fig.~\ref{fig:hists}c, the histogram shows the distribution of the observed amplitudes. 
%For the debiased amplitudes, we followed the method given by \cite{Binzel1992}. 
We do not apply any de-bias procedure to the amplitudes as spin axes of Trojans appear to prefer perpendicular orientations to the plane of the Solar System instead of a uniform distribution \citep{Bowell-2014}. Where both minimum and maximum values were listed in the LCDB, we used the average value for this figure.
Most of the asteroids have amplitudes similar to the median value, but multiple asteroids fall into the 0.6--1.0 mag range too. Asteroids in this range could be either strongly elongated, or binary objects. 

\begin{figure}
\includegraphics[width=0.95\columnwidth]{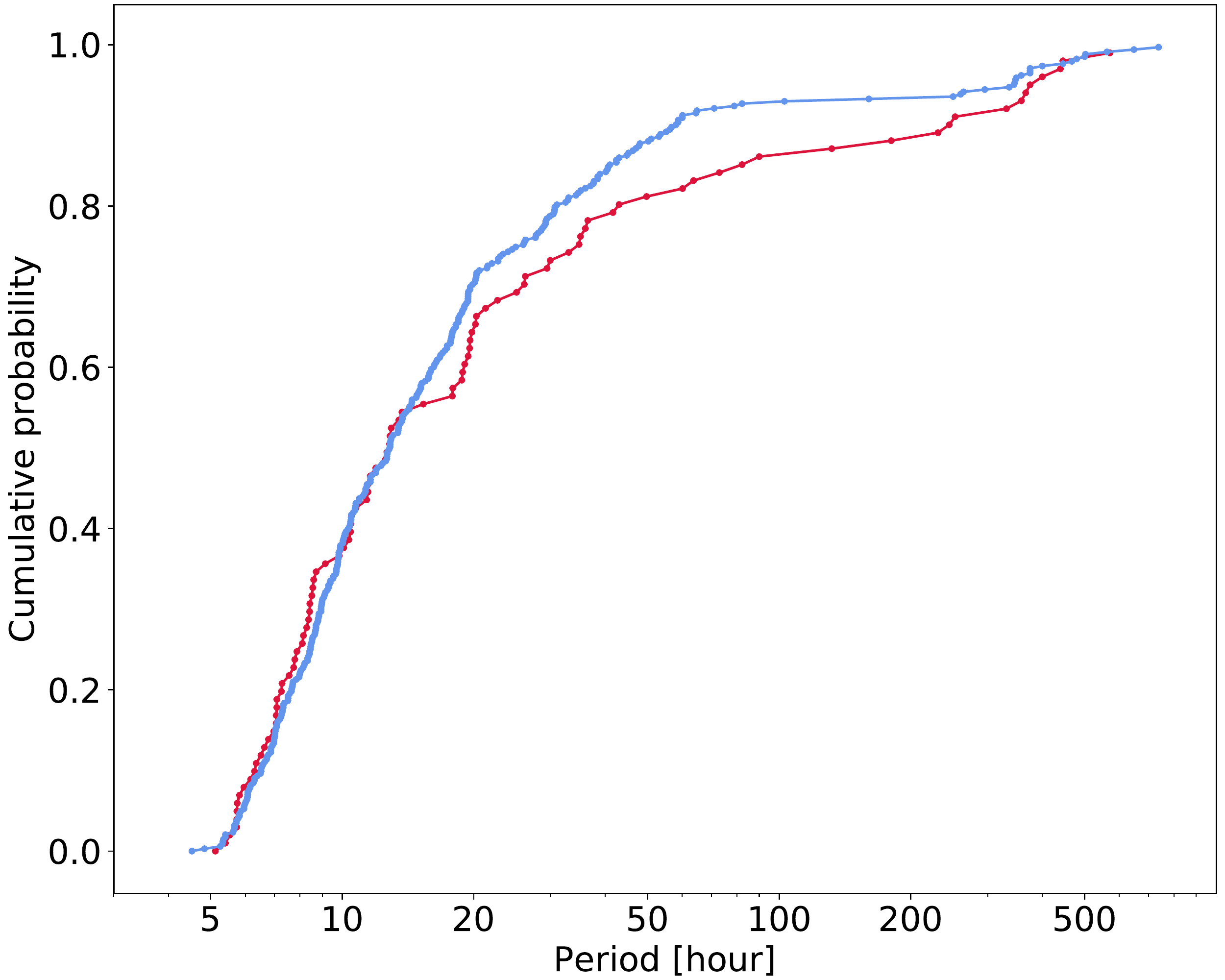}
\caption{Cumulative distribution of Trojan periods from LCDB (blue) and from K2 (red).}
\label{fig:ks-plot}
\end{figure}

In Fig.~\ref{fig:ks-plot}, we plot the cumulative distribution of Trojan rotation periods obtained from the LCDB (blue) and from the K2 observations (red). Above $\sim15$ hr the slope of the cumulative distribution of the LCDB sample remains steeper than that of the K2 sample, and the difference becomes most pronounced at the long-period wing of the distribution. This means Trojans with longer periods are less represented in the LCDB than in the K2 sample.

\begin{table}
\centering
\begin{tabular}{ccccc}
\hline
\hline
     &  MB   & MB$_{TESS}$ & JT & Hilda \\
\hline
        N              & 13072  & 7874  & 101   & 112       \\
f$_m$ [cycle d$^{-1}$] & 3.53   & 1.49  & 1.90  & 1.26    \\  
P$_m$ [h]              & 6.79   & 16.06 & 12.65 & 19.02   \\
\hline
N$_{f}$               & 1930   & 733      & 0   & 2   \\
r$_{f}$ [\%]          & 14.8   & 7.4      & 0   & 1.8     \\
\hline
N$_{s}$               & 1445   & 3463   & 26    &  43   \\ 
r$_{s}$ [\%]          & 11.1   & 34.9   & 25.7  &  38.9 \\
\hline
N$_{vs}$              & 488  & 1390    & 13     & 20      \\ 
r$_{vs}$ [\%]         & 3.7  & 14.0    & 12.9   & 17.85  \\
\hline
\end{tabular}
\caption{Summary table of median rotation rates (f$_m$, and the corresponding period P$_m$), and the number of slow and fast rotating asteroids in the main belt, as obtained from data in the LCDB (MB), main belt asteroid data from the TESS DR1 (MB$_{TESS}$) \citep{Pal2020ApJS}, K2 Jovian trojans (JT, this work), and K2 Hilda asteroids \citep{Szabo2020ApJS}. We defined fast rotators (subscript 'f') as f\,$\geq$\,7\,d$^{-1}$ (P\,$\leq$\,3.43\,h), slow rotators ('s') as f\,$\leq$\,0.8\,d$^{-1}$ (P\,$\geq$\,30\,h) and very slow rotators ('vs') as f\,$\leq$\,0.24\,d$^{-1}$ (P\,$\geq$\,100\,h), following \citet{Pravec2007}.}
\label{table:fstat}
\end{table}

\begin{figure}
\includegraphics[width=0.95\columnwidth]{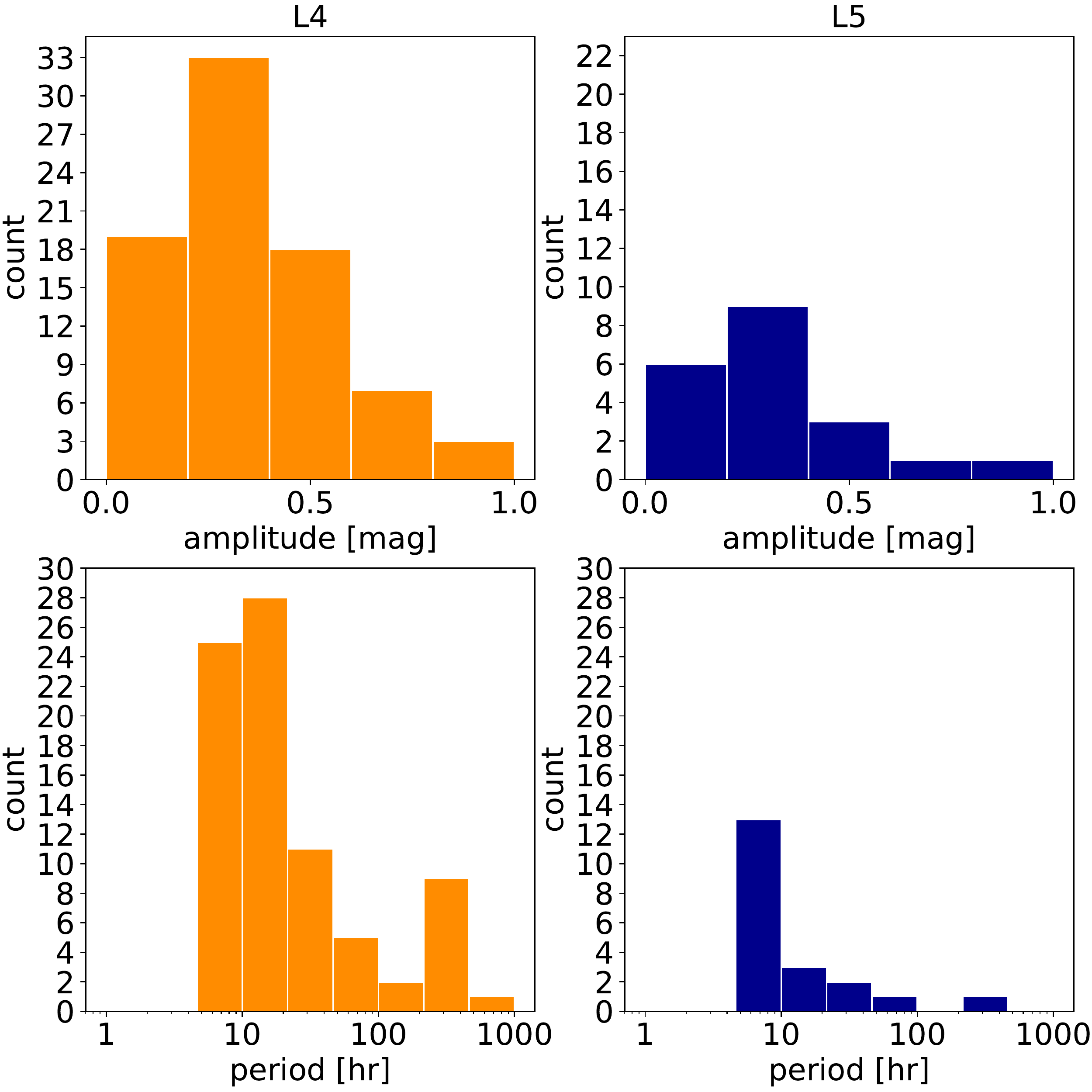}
\caption{Comparison of the distributions of the L4 and L5 nodes of Jovian Trojans.}
\label{fig:L4-L5}
\end{figure}

We also compared the distributions of the swarms in Fig.~\ref{fig:L4-L5} and conclude that there is no significant difference between them, which is consistent with previous statistical results \citep{Slyusarev2018}.

\begin{figure}
\includegraphics[width=0.98\columnwidth]{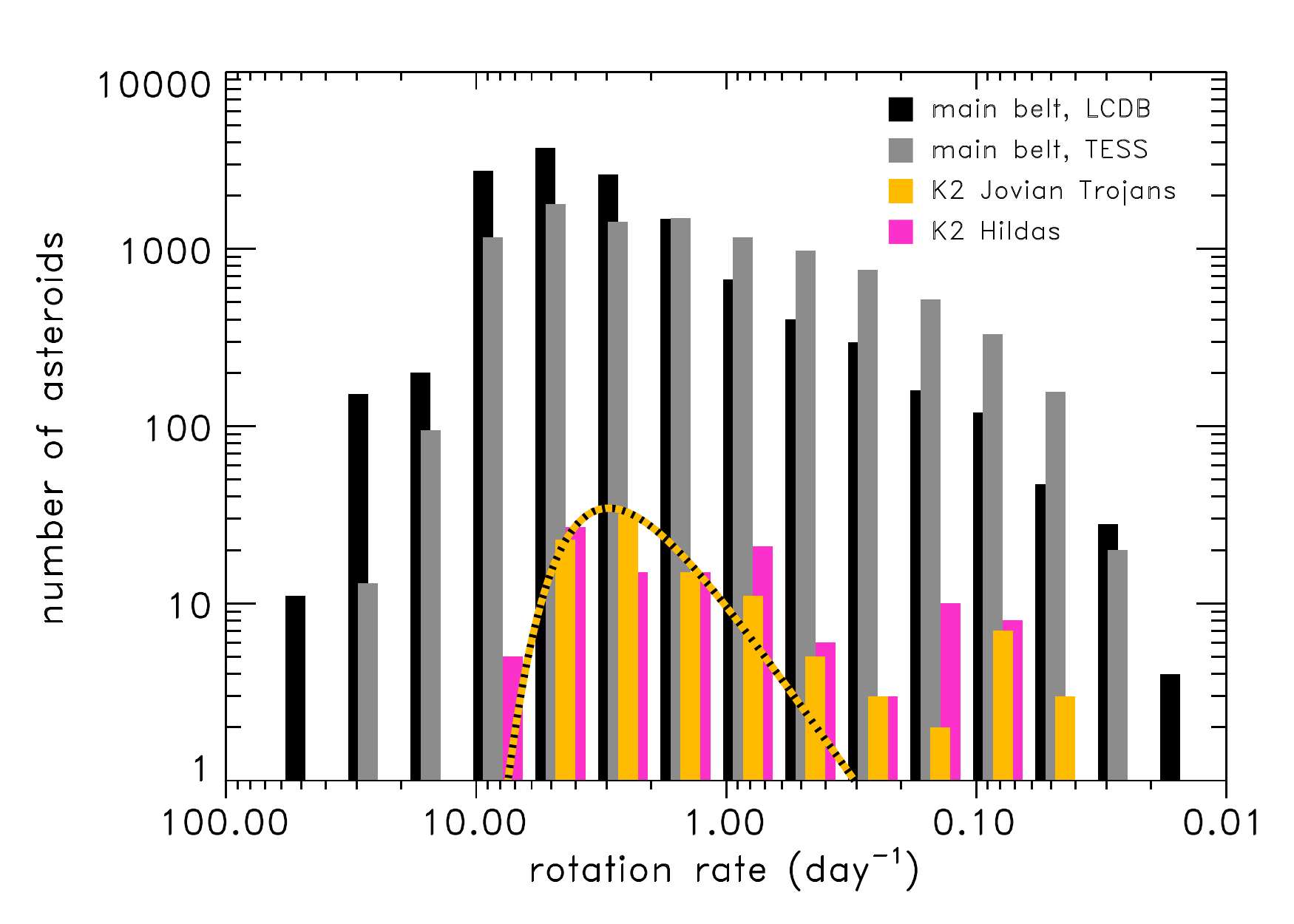}
\caption{Rotational frequency distribution of Jovian trojan asteroids from the K2 mission (orange bars) compared with those of other asteroid samples (black: main belt asteroids from the LCDB; gray: main belt asteroids form the TESS DR1 sample, \citealt{Pal2020ApJS}; magenta: K2 Hilda asteroids, \citealt{Szabo2020ApJS}). The black-orange dashed curve represent the Maxwellian fit to the K2 Jovian trojan sample.}
\label{fig:period-barplot}
\end{figure}

If we compare our K2 Jovian trojan sample to the properties of other types of asteroids derived from \textit{Kepler} \citep{Szabo2020ApJS} and TESS \citep{Pal2020ApJS} data, we find that slow rotators exist elsewhere, too. 
It is also seen that ground-based observations underestimated the number of slow rotators in the Main Belt. In Fig.~\ref{fig:period-barplot} we plot the rotation rate distributions of these populations. In the cases of Trojans and Hildas, we can see a possible dichotomy in rotational periods. We fitted a Maxwellian distribution, which describes a collisionally evolved sample, to the Trojan sample. While the short-period group can be fitted reasonably well with this profile, the long-period group clearly falls outside the distribution. We also computed the ratio of fast, slow and very slow rotators among the four samples mentioned above, in Table~\ref{table:fstat}. This, again, shows that there are essentially no fast rotators among the Trojans and Hildas, whereas a large portion of the sample rotates slowly or very slowly. Moreover, with the TESS observations at hand, the rate of slow and very slow rotators among the main-belt asteroids is comparable to that of the farther groups of Hildas and Trojans.

\begin{figure}
\includegraphics[width=0.95\columnwidth]{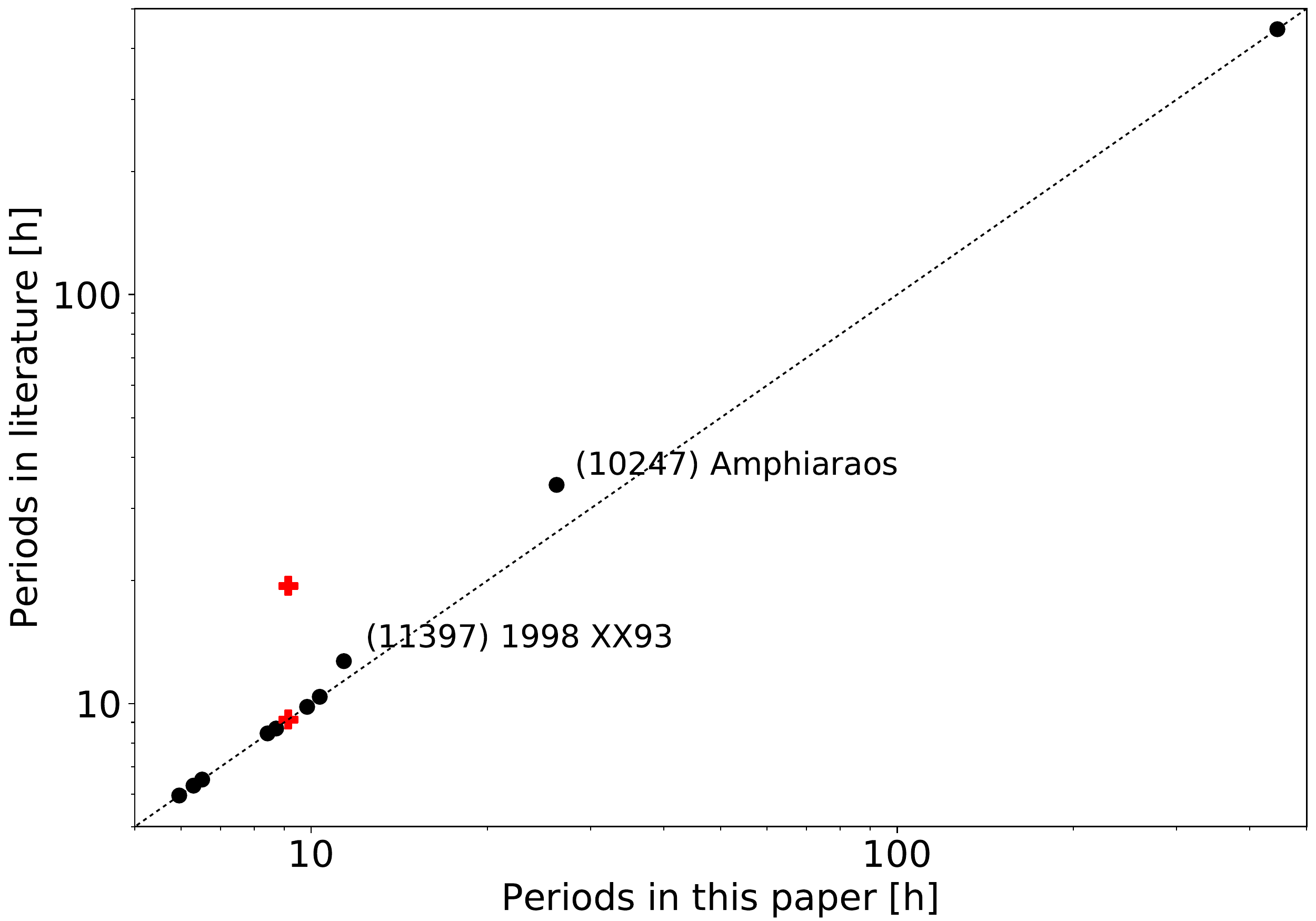}
\caption{Comparison of periods in the literature and this paper. Red plus signs represent (5638) Deikoon, that have two different rotational period in the literature.}
\label{fig:lit_P_comp}
\end{figure}

We compare our period determinations with existing literature values in Fig.~\ref{fig:lit_P_comp}. It shows that above 10 hr the previous rotational periods are significantly different from ours, except for the largest value, which belongs to (11351) Leucus. Leucus is an important asteroid as it will be one of the targets of the $Lucy$ mission, hence previous studies measured its period quite accurately \citep{Buie2018}. In another three long-period cases (5638, 11397, 10247) we have fully covered phased light curves with precise periods, which shows the advantage of the uninterrupted K2 measurements compared to ground-based observations, where the changing circumstances and daily aliases make it more difficult to achieve high accuracy. 

In the case of (5638) Deikoon, the current default rotation period value in the LCDB is $19.396\pm0.011$ hr, based on the observations of \citet{Molnar-2008} who constructed a complex light curve with multiple maxima and minima. Later, \citet{Mottola-2011} also detected a complex, four-peaked light curve, but with a rotation period of $9.137\pm0.003$ hr. This latter value is very close to the $9.146\pm0.029$ hr period we calculated from the K2 data, and the shape of the light curve is also similar to our photometry, indicating that this shorter value is the correct rotation period for the asteroid.

\subsection{Period-amplitude diagram}
\begin{figure}
\includegraphics[width=0.98\columnwidth]{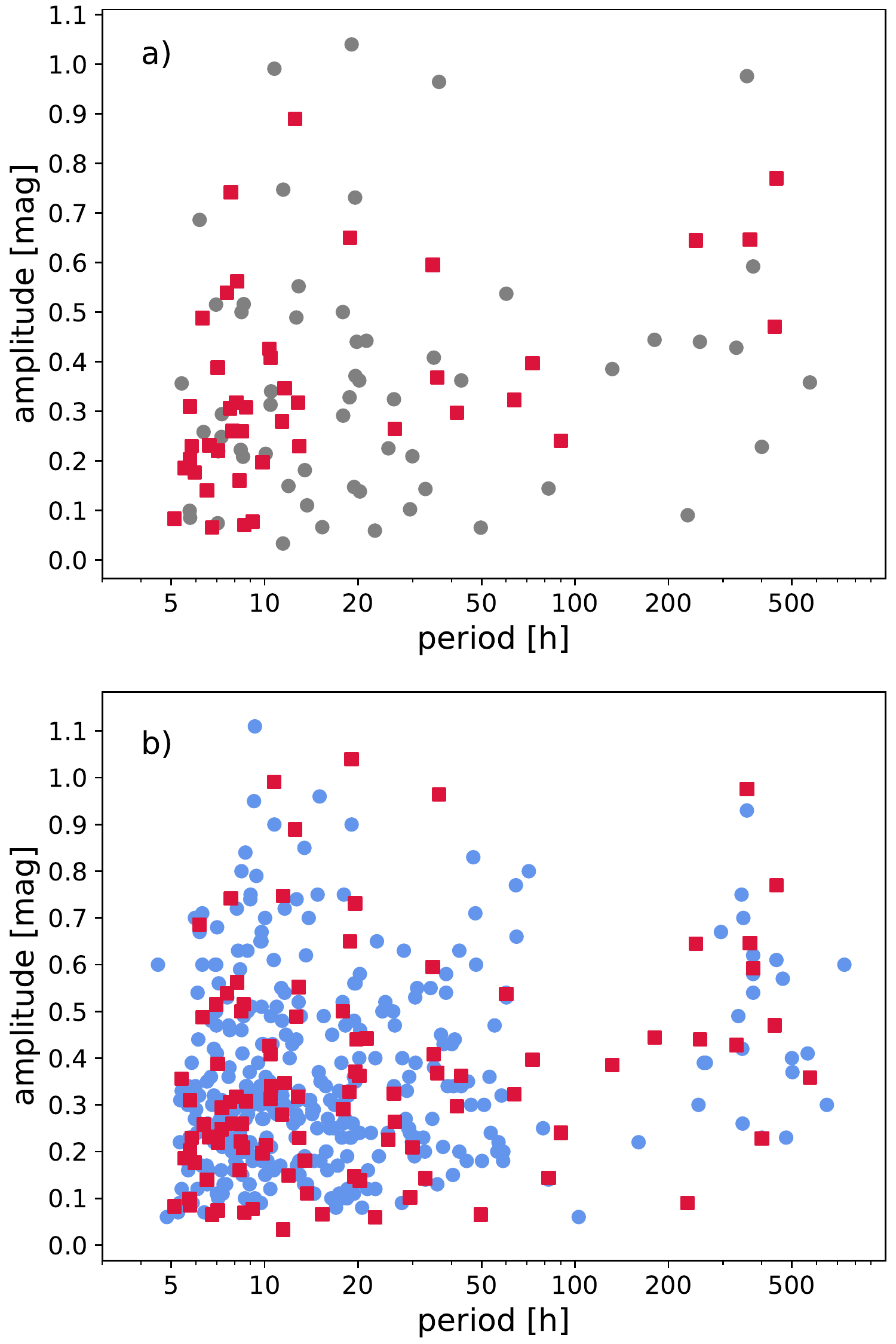}
\caption{Period-amplitude distributions of Jovian trojan asteroids. Panel a) shows the observed K2 amplitudes, where grey dots are Trojans from C6 and red squares from our sample. In panel b), red squares are the overall K2 Trojan sample, and blue dots are Trojans from LCDB, using average amplitudes if minimum and maximum amplitudes were both listed in the database.}
\label{fig:per_amp}
\end{figure}

Panels of Fig.~\ref{fig:per_amp} show the period-amplitude distributions of Jovian trojan asteroids. In panel (b) and (c) we used bias-corrected amplitudes. Fig.~\ref{fig:per_amp}c shows Trojans from LCDB (blue dots) with Trojans from K2 (red squares). We can notice the same period dichotomy that was evident in Fig.~\ref{fig:hists} too, with a smaller group clustering at periods above 100 hr. Although the de-biased amplitudes group at lower values, there still seems to be gap in large amplitudes ($>0.4$~mag) between 50--200 hr. The sample size of the large-amplitude group is too small to probe the existence of a gap statistically, but we note that a similar feature can be seen among the Hilda observations too \citep{Szabo2020ApJS}.

Furthermore, the LCDB sample is clearly more populous in the short-period range: this was also evident already from the period histogram in Fig.~\ref{fig:hists}. These distributions clearly show that further, extended and precise photometry is key to determine the range of Trojan rotation amplitudes and rates that could possibly extend to the 1000-hour range.

\section{Analysis}
\label{sect:analysis}

\subsection{Spin barrier and critical density}
\label{sec:spinbarrier}
Minor planets have a rotational limit called the spin barrier, which value is the smallest possible rotational period that a given type of asteroid can have without flying apart due to its centrifugal acceleration. The critical period can be calculated approximately as 

\begin{equation}
 P_c \approx 3.3 \cdot \sqrt{\frac{(1+A)}{\rho_c}}    
 \label{eq:crit}
\end{equation}

where $P_c$ is expressed in hours, $A$ is the amplitude of the light curve, and $\rho_c$ is the critical density in g\,cm$^{-3}$, which is a lower limit estimate of the asteroid's bulk density \citep{Pravec2000}.

\begin{figure}
    \centering
    \includegraphics[width=0.5\textwidth]{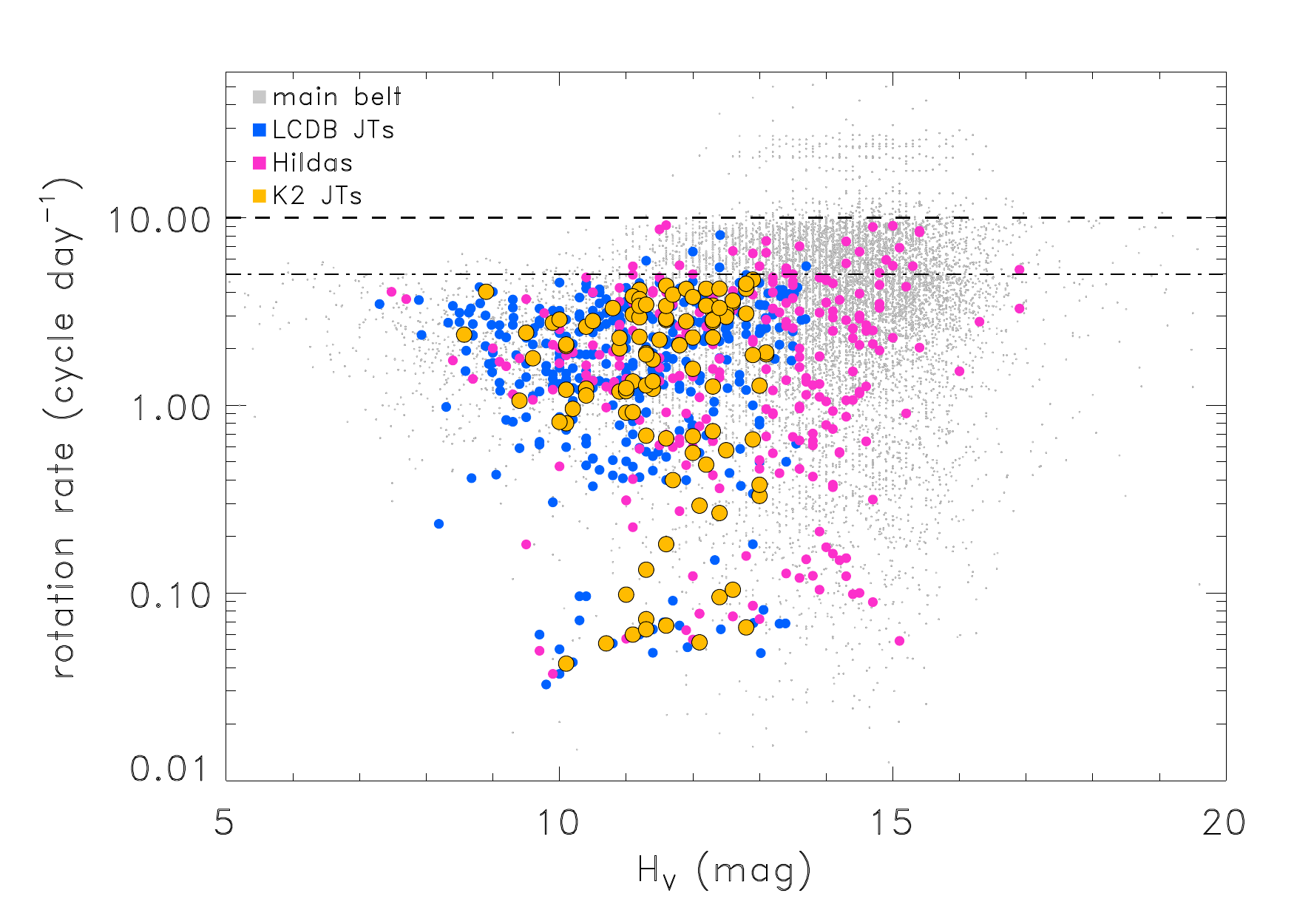}
    \caption{ Spin rate versus $V$-band absolute magnitude of main belt, Hilda, and Jovian trojan asteroids. The horizontal dashed and dash-dotted lines represent the spin rates of f\,=\,10 and 5 d$^{-1}$ or the rotation periods of P\,=\,2.4 and 4.8\,h. These correspond to critical densities of $\sim$2 and $\sim$0.5~g\,cm$^{-3}$.}
    \label{fig:f_vs_Hv}
\end{figure}

In Fig.~\ref{fig:f_vs_Hv} we present the distribution of rotation rate versus the absolute magnitude of the asteroids for our Jovian trojan sample, as well as for main belt asteroids and Hildas. For main belt asteroids the 'spin barrier' at $\sim$10 d$^{-1}$ (dashed horizontal line) corresponds a critical density of $\sim$2 g\,cm$^{-3}$ calculated from Eq.~\ref{eq:crit}. Although there are a few Jovian trojans in the LCDB that have relatively short rotation periods, the bulk of Jovian trojans -- most targets in the LCDB (blue dots), and all targets in the K2 sample (orange symbols) -- have a spin rate f\,$\lesssim$\,5 d$^{-1}$ or rotation period P\,$\gtrsim$\,4.8\,h. This corresponds to a critical density of 0.5 g\,cm$^{-3}$, significantly lower than in the main belt. 

\begin{figure}
\includegraphics[width=1.0\columnwidth]{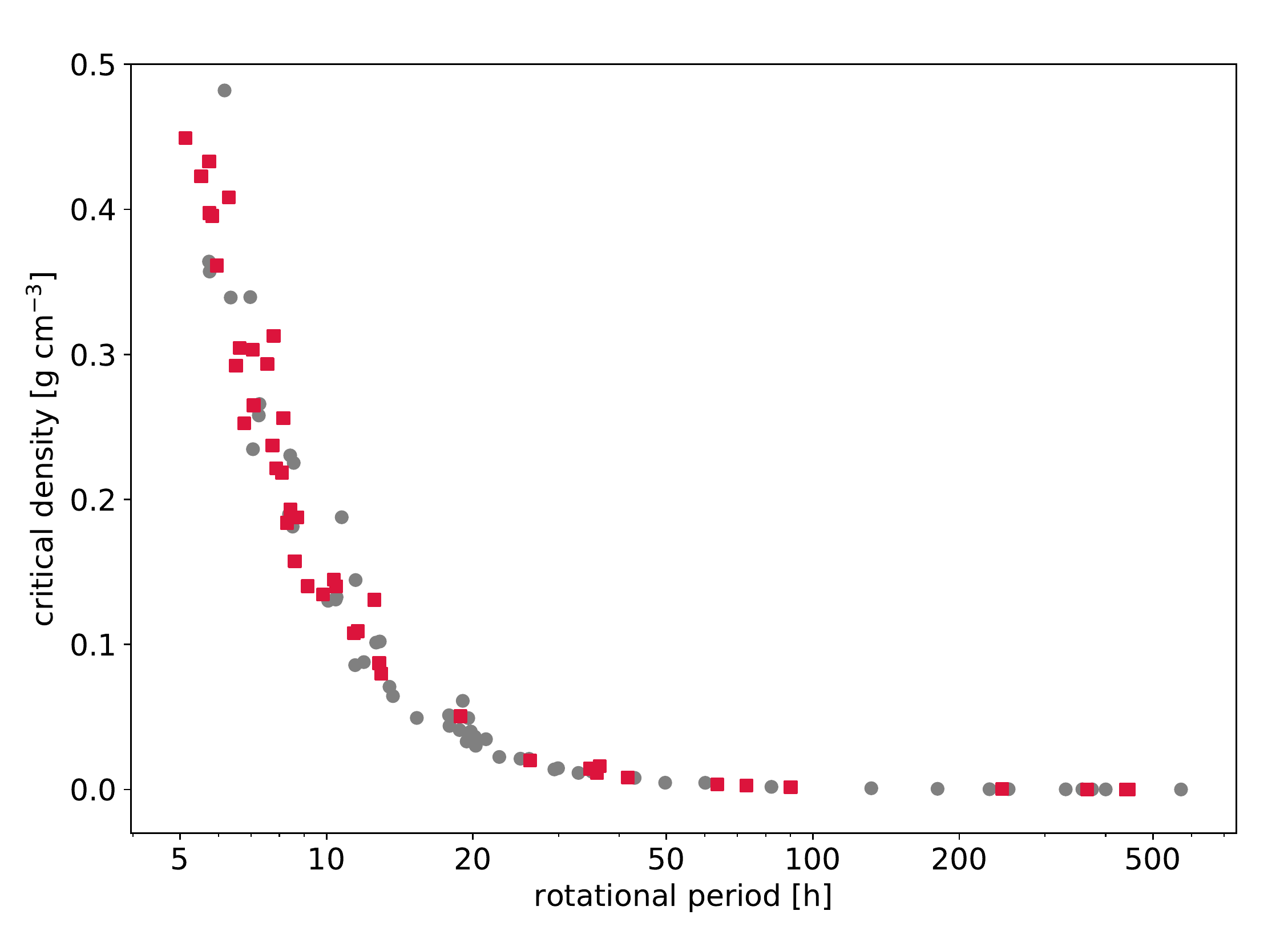}
\caption{Critical densities derived from the overall sample of K2 Jovian trojans as a function of rotational period. Grey points are from C6, red squares are from our sample.}
\label{fig:crit_den}
\end{figure}

Using Eq.~\ref{eq:crit} we calculated the critical densities each of the asteroids in our sample. In Fig.~\ref{fig:crit_den} we present these critical densities complemented with Jovian trojans from C6 as a function of rotation periods. It shows that there are no points below $\sim$ 5.1 hr and above $\sim$ 0.5 g$\,$cm$^{-3}$. 

If our sample is unbiased these values estimate the lower limit of rotational period and the upper limit of the critical density of Jovian trojans. We can rule out most observational biases safely: we detect variations in all but one of the 101 Trojans. There is a small chance that some of the long variations are caused by stroboscopic sampling of a rotation period close to the sampling frequency of \textit{Kepler}. However, as K2 data for main-belt asteroids show, we would be able to detect rotation rates up to twice as short as the fastest Trojan in our sample before reaching the Nyquist limit \citep{Molnar2018}. There are no Trojans with a measured period in the LCDB that would rotate fast enough to suffer from stroboscopic sampling issues if observed with the \textit{Kepler} cadence either. Therefore we can conclude that we observe the true rotation limit of the sample.

It is also possible that only Trojans below the faint limit of the K2 observations ($H_V \gtrsim 13$ mag, or under 10--20 km diameter) rotate faster. At this distance the YORP effect may not be sufficient to spin large objects up (or down). However, as our timescale calculations in the Appendix indicate, collisions can spin up Trojans much more effectively than the YORP effect. Indeed, as Fig.~\ref{fig:period-barplot} indicates, the fast-rotating group of Trojans appear to follow a Maxwellian distribution, suggesting collisional evolution.

Based on the Maxwellian fit of the short-period group in Fig.~\ref{fig:period-barplot} we calculated the expected number of asteroids with periods between 3--5 hr. In a sample with 100 elements, we predict 10--30 asteroids with periods below 5~hr, while below 3~hr the expected number drops to only 0--4 objects. We also calculated the probability that zero objects have periods below a given period in that sample. We find that at 3~hr there is a 38\% chance of this. The probability increases steeply below 3 hr, it is more than 50\% at 2.9 hr already. We conclude that in the collisionally evolving model, we would expect to find multiple objects between 3 and 5 hr in a sample of 100 asteroids. Neither the Maxwellian distribution, nor the YORP effect indicate that we would not detect asteroids with rotational period below 5 hr, which implies there should be another physical reason to explain the cutoff the short period end of the distribution. We interpret the observed cutoff at $\approx 5$ hr as the manifestation of the spin barrier, which suggests that bulk densities of Trojans reach lower values than those of main belt asteroids.

The critical density limits that we and other previous calculations obtained for Jovian trojans are also significantly different from that of the main belt, but very similar to the TNO population or the nucleus of comets. 
This value is expected from icy bodies with notable porosity, and it is similar to the bulk densities found for trans-Neptunian objects in the D\,$\le$\,500\,km size range \citep[see e.g.][]{Grundy2019}. It is also in agreement with color and albedo measurements \citep{Szabo2007}.

\subsection{Rate of binarity}

We estimated the rate of binarity of our sample in two different ways.
First, we used two features that can indicate a binary asteroid, depending on the photometric properties of the asteroids: a.) at least one phase angle, the amplitude should exceed 0.9$^{\mathrm{m}}$ \citep{Leone1984}, and b.) the rotational period should be greater than the triple of the median value \citep{Sonnett2015}. Here we approximate the limit of slow rotation to be 30 hr \citep{Pravec2007}. In our sample there are 1 (200037) and 11 (11351, 13062, 18071, 26486, 31819, 96337, 99306, 111113, 175471, 200037, 296787) asteroids that fulfill these amplitude and period requirements, respectively. This leads to an estimated $\sim$ 25\% rate of binaries among our sample, which is consistent with previous calculations \citep{Szabo2017}. However, this is estimate is different from constraining the true binary rate, because some objects may produce the expected features through other mechanisms, and some binaries will fall outside these selection bounds.

\begin{figure*}
\hbox{\includegraphics[width=0.35\textwidth]{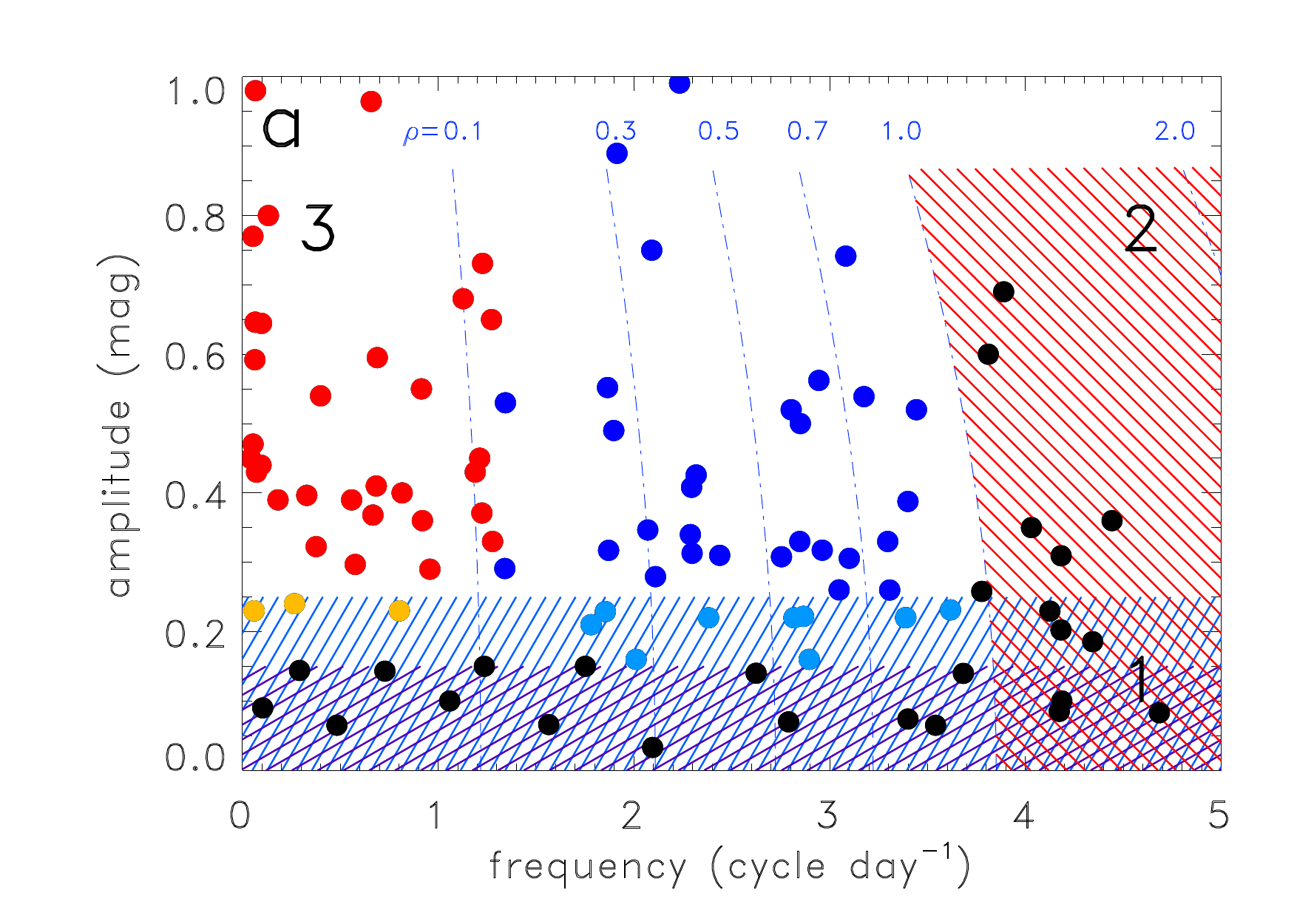}
\hskip -0.6cm
\includegraphics[width=0.35\textwidth]{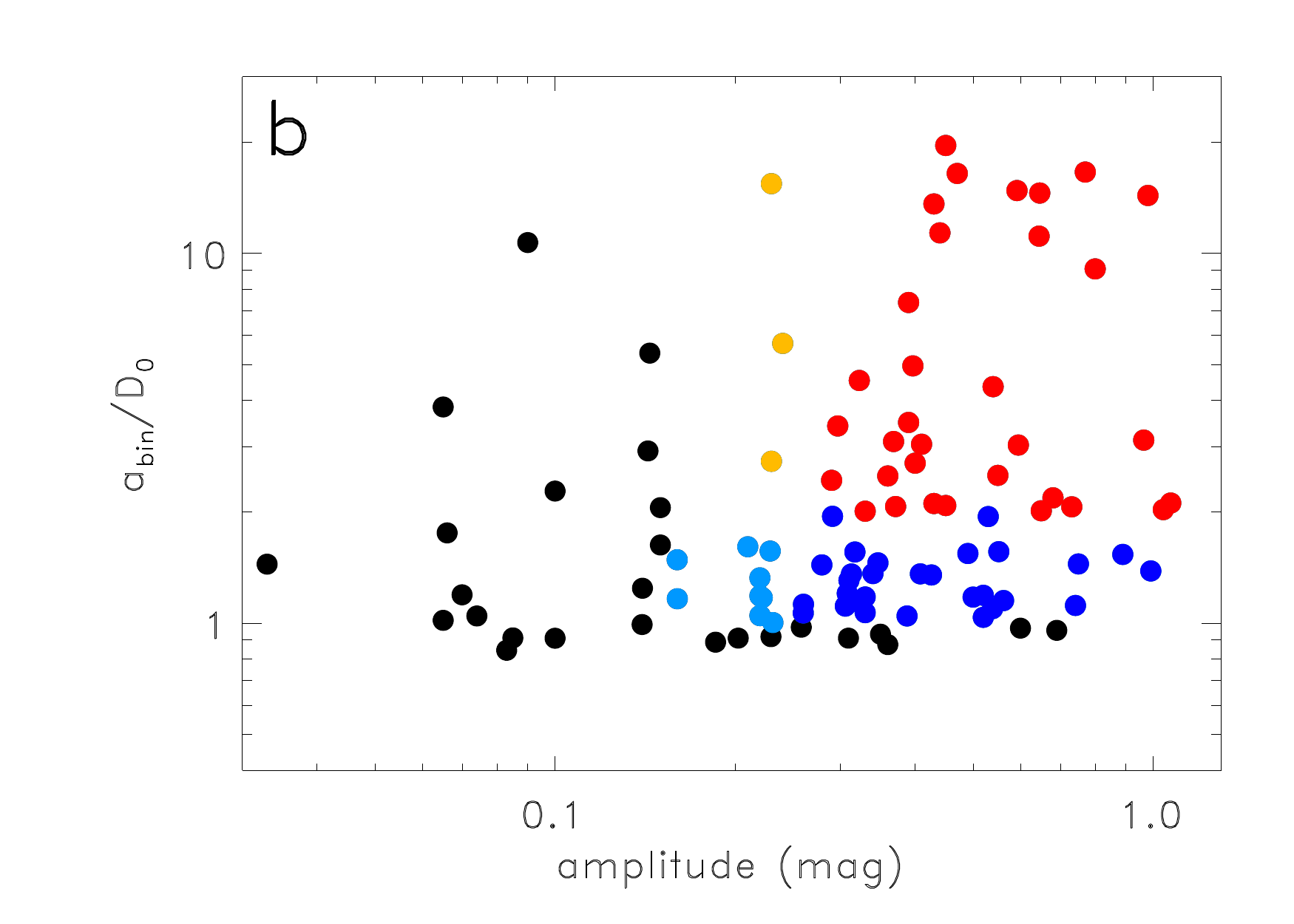}
\hskip -0.6cm
\includegraphics[width=0.35\textwidth]{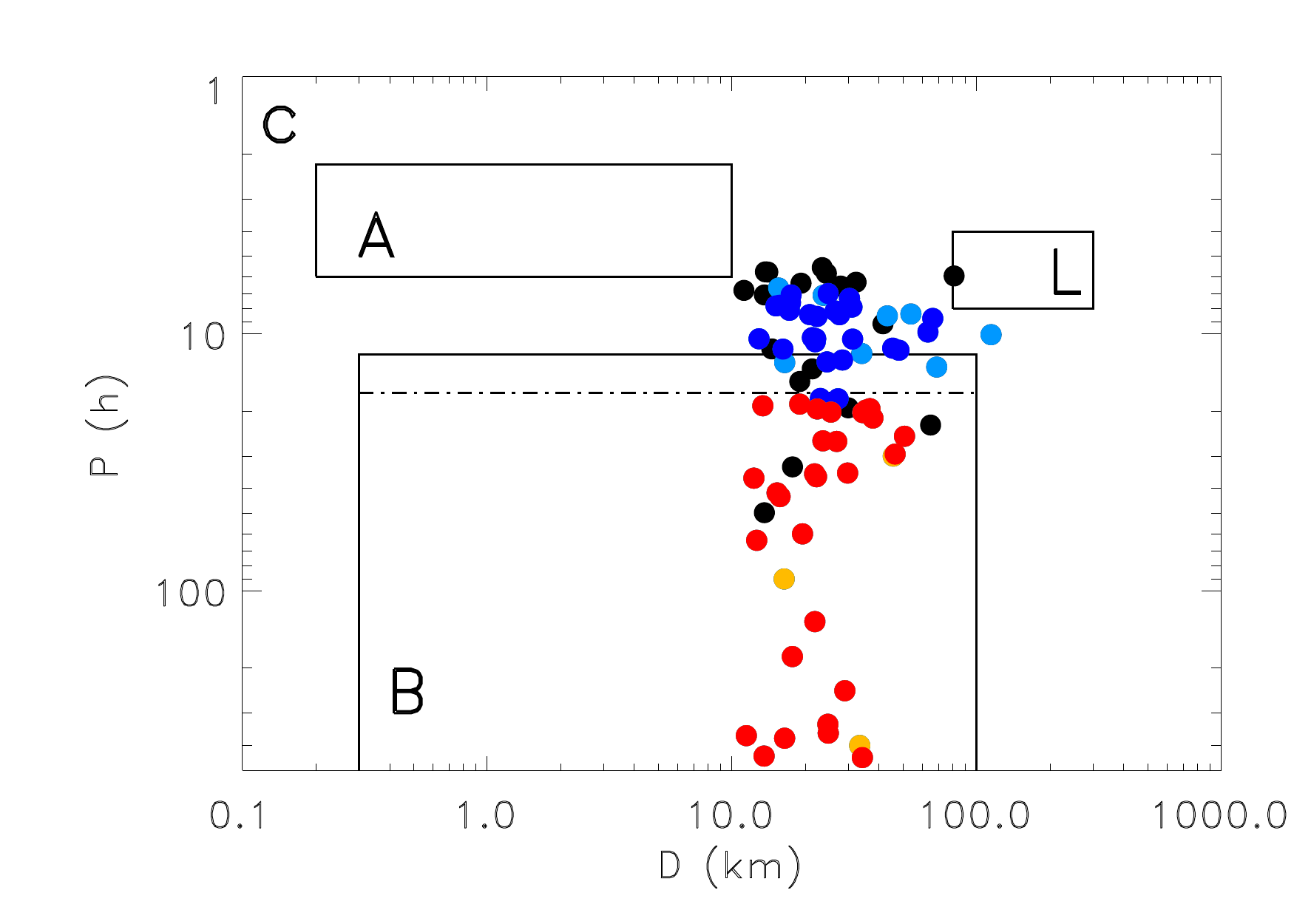}}
\caption{(a) Light curve amplitude versus frequency of Jovian trojans in this paper and in \citet{Szabo2017}. Blue dash-dotted curves represent the rotational frequencies and light curve amplitudes corresponding to the rotation of a strengthless body (Jacobi ellipsoid), of a specific density (shown at the top in g\,cm$^{-3}$ units). In the blue and purple shaded areas (below $\Delta$m\,$\leq$\,0.25 or 0.15\,mag, region~1) light curves can be explained either by albedo variations, deformed shape or binarity. Targets in the red shaded area (region~2), right of the $\rho$\,=\,1.0 g\,cm$^{-3}$ curve (or any other constant density chosen), could be elongated due to rotation. Objects in region~3 should have densities notably below $\rho$\,=\,1.0 g\,cm$^{-3}$ in order to be elongated from rotation.
(b) Ratio of binary orbit to binary diameter ($a_{bin}/D_0$) versus the maximum light curve amplitude ($A_{max}$). 
(c) Rotation period versus deduced diameter. The black boxes mark the regions~A, B and L identified by \citet{Pravec2007}. 
Colour symbols in all subfigures correspond to the following: 
light blue -- 1\,$\leq$\,$a_{bin}/D_0$\,$\leq$\,2 and 0.15\,$\leq$\,$A_{max}$\,$\leq$\,0.25\,mag;
dark blue -- 1\,$\leq$\,$a_{bin}/D_0$\,$\leq$\,2 and $A_{max}$\,$\geq$\,0.25\,mag;
orange -- $a_{bin}/D_0$\,$\geq$\,2 and 0.15\,$\leq$\,$A_{max}$\,$\leq$\,0.25\,mag;
red -- $a_{bin}/D_0$\,$\geq$\,2 and $A_{max}$\,$\geq$\,0.25\,mag. The dash-dotted line marks the separation between the red and dark blue samples.}
\label{fig:binarity}
\end{figure*}

To get some hints on the reason behind their light curve variations, minor bodies are often classified based on their spin frequencies and light curve amplitudes:  
\citet{Leone1984} and \citet{Sheppard2004} identified three main zones on the light curve amplitude versus rotational frequency plane (see Fig.~\ref{fig:binarity}a), re-evaluated by \citet{Thirouin2010} and \citet{Benecchi2013}. Light curve variations of objects with small amplitudes ($\Delta m$\,$\leq$\,0.25\,mag or 0.15\,mag) can either be caused by albedo and shape features or can as well be binaries where the orbital plane is nearly face-on from our vantage point (region~1 in Fig.~\ref{fig:binarity}a). If the rotational equilibrium of a strengthless body is considered and approximated by a Jacobi ellipsoid, constant density curves can be drawn (blue dash-dotted curves in Fig.~\ref{fig:binarity}a), assuming $\vartheta$\,=\,$\pi/2$ aspect angle, i.e. equator-on viewing geometry and maximum light curve amplitude. Objects to the right of a curve of a constant density (e.g. 1.0~g\,cm$^{-3}$, region~2) are likely rotating single bodies, if their rotational speed is below the breakup limit. The rotation of the objects to the left (region~3) is too slow to cause elongation and a corresponding rotational light curve. For these objects the light curves are often explained by binarity \citep[e.g.][]{Leone1984,Sheppard2004}. 

As panel (a) of Fig.~\ref{fig:binarity} indicates, densities do not exceed 1.5 g\,cm$^{-3}$ when assuming a strenghtless body model. This is in agreement with the spin barrier result presented before in Sect.~\ref{sec:spinbarrier}, indicating that the Jovian trojans generally have low densities, suggesting icy and/or porous composition. 

We can also characterise the possibly binary nature of a specific system following \cite{Marton2020}. We use the estimated 'separation', $a_{bin}$, the semi-major axis of the orbit of the potential binary, assuming that the binary has two equally sized, spherical (of diameter $D_0$) and equal-mass components, and a uniform density of 1 g\,cm$^{-3}$. Effective diameters are obtained from the {\it NEOWISE Derived Diameters and Albedos of Solar System Small Bodies\footnote{\url{https://sbn.psi.edu/pds/resource/neowisediam.html},\\ \url{https://irsa.ipac.caltech.edu/cgi-bin/Gator/nph-dd}}} database \citep{Mainzer2019}. If no geometric albedo was available then we used the median value of the known geometric albedos, $p_V$\,=\,0.079, and calculated the effective diameter from the absolute magnitude \citep[see][for details]{Marton2020}.  The $a_{bin}/D_0$\,=\,1 case is a classical contact binary. While formally the minimum requirement for a binary in this scheme is $a_{bin}/D_0$\,$\geq$\,1 for two spherical objects, contact or semi-contact binaries that are modelled, e.g., as Roche systems have elongated components and thus require larger separations (hence larger $a_{bin}/D_0$) \citep[see e.g.][]{Lacerda2007}. Therefore we consider $a_{bin}/D_0$\,$\geq$\,2 as a more suitable limit between contact and detached binaries. In our sample we found 10 systems (namely 11351, 13062, 18071, 26486, 31819, 96337, 111113, 175471, 200037, 296787) consistent with our estimation from photometric properties, and in the full K2 Jovian trojan sample 32 systems (11351, 13062, 18071, 26486, 31819, 111113, 296787, 316484, 96337, 10247, 175471, 200037, 3801, 4138, 5028, 5123, 5244, 5436, 9807, 10989, 13331, 13372, 14791, 15529, 22056, 24357, 35363, 39289, 63239, 65210, 65223, 129602) for which both the separation, $a_{bin}/D_0$\,$\geq$\,2 and the amplitude, $A_{max}$\,$\geq$\,0.25\,mag, are large enough to classify them as potential detached binary systems (red symbols in Fig.~\ref{fig:binarity}). However, we cannot rule out that some of these might be members of dissociated binaries: this seems likely for (11351) Leucus for example, as we discuss it in Sect.~\ref{sec:leucus}.

Among the main-belt binary asteroids \citet{Pravec2007} identified three main groups, based on their angular momentum contents (regions A, B and L in Fig.~\ref{fig:binarity}c). In group L we find large asteroids with small satellites and systems, whereas group~A is composed of small binaries with fast rotation, likely due to the YORP effect \citep{Rubincam-2000}. Our sample contains no small objects that would clearly fall into group~A, but we identified one member of the L~group, (5144) Achates. Currently we have no observational indication of Achates being a binary system. Group~B represents double synchronous (spin-locked) systems. While Jovian trojans may have physical properties different from that of main belt asteroids, Fig.~\ref{fig:binarity}c shows that all of our Jovian trojans identified as binary candidates based on their $a_{min}/D_0$ ratio and light curve characteristics---and assuming spin locking---are in region~B, where double synchronous systems can be found in the main belt.  

The yellow and light blue points in Fig.~\ref{fig:binarity} are close to the maximum photometric amplitude a strengthless ellipsoid can achieve, and only differ by the separation parameter. These are considered edge cases whose classification will need further observations at different phase angles and orbital geometries.

Dark blue points represent asteroids where the strengthless model cannot account for the photometric amplitude but the estimated separation is too small for a detached binary. The objects in this region could either be single objects with large surface features, elongated, non-axisymmetric objects, or could also be contact/collapsed binary asteroids. Examples for these have been observed directly by various space missions. Photographs taken by the NEAR-Shoemaker probe showed that the shape of the rubble pile asteroid (253) Mathilde is dominated by five deep giant craters \citep{Thomas1999}. The probe also confirmed that the near-Earth asteroid (433) Eros rotates about one of its minor axes, as indicated earlier radar and photometric results \citep{Ostro1990,Miller2002}. These properties can increase the photometric amplitudes. 

Finally, multiple contact binary objects, including near-Earth asteroid (4179) Toutatis, 67P/Churyumov-Gerasimenko and (486958) Arrokoth are thought to have formed through low-speed impacts of a binary pair, or---in the case of 67P---through low-energy sub-catastrophic collisions that leave behind a bi-lobate remnant \citep{Hu2018MNRAS,Jutzi2017,Grishin2020}. If Trojans originate from the outer Solar System, we could expect collapsed binaries similar to (486958) Arrokoth among them as well.

\subsection{The incidence of very slow rotators}

\citet{nesvorny2020} estimated a rate $\sim$15\% of very slow rotators (P\,$\ge$\,100\,h) in the C6 sample. In our sample we identified four new slow rotators: (11351) Leucus, (13062) Podarkes, (26486) 2000 AQ231 and (96337) 1997 LG2. With those in the C6 sample this means altogether 13 very slowly rotating targets, $\sim$13\%, of the total Jovian trojan K2 sample. 

We estimated the YORP timescale $\tau_{\rm YORP}$ for all of the very slow rotators, as described in the Appendix.  %following \citet{nesvorny2020}. 
For all these targets---which are $D$\,$\ge$\,10\,km---we find $\tau_{\rm YORP}$\,$\ge$\,10$^9$\,yr. %, the lower limit belonging to asteroid (96337) 1997~LG$_2$. 
On the other hand, the timescale to notably change the rotational state of an asteroid via collisions is 2-3 orders of magnitudes shorter than $\tau_{\rm YORP}$ for Jovian trojans ($\sim$20--66\,Myr, see Appendix), suggesting that collisions can effectively override the effect of slow spin-down though YORP. This therefore favours other mechanisms, such as tidal synchronization, for the origin of the slow rotation. \citet{nesvorny2020} estimated that Jovian trojan binaries with $a_{bin}/D_0$\,$\lesssim$15--20 should become synchronous by the age of the solar system. This corresponds to periods up to 400--750\,hr, i.e., tides are capable of producing very long rotation periods.

\subsection{Notable targets}
We inspected the light curves of the targets individually. Three asteroids from the sample were notable for various reasons. In this section we discuss their photometric results in more detail. 

\begin{figure}
\includegraphics[width=0.95\columnwidth]{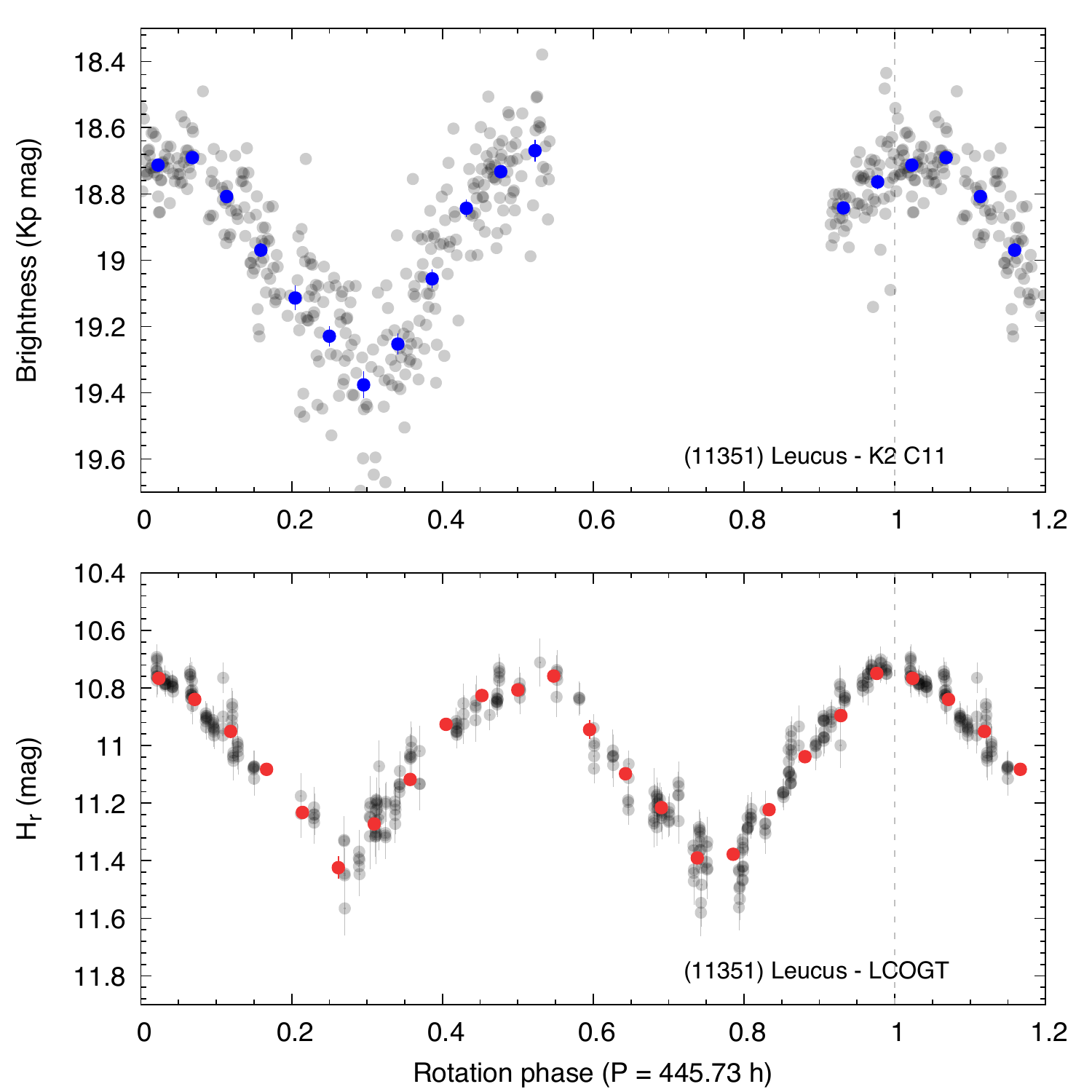}
\caption{Phase-folded light curves of (11351) Leucus. Top: the K2 data; bottom: ground-based photometry collected by \citet{Buie2018}. Blue points are individual measurements, black points are phase-binned values.}
\label{fig:Leucus}
\end{figure}

\subsubsection{(11351) Leucus}
\label{sec:leucus}
Perhaps the most famous asteroid among the ones presented in this study is (11351) Leucus, which was chosen as one of the flyby targets of the \textit{Lucy} mission \citep{Levison2016,Levison2019}. Leucus belongs to the extremely slow rotators, as first reported by \citet{French2013}. A more detailed study based on more extensive ground-based observations led to a refined period of $P=445.732$ hr  \citep{Buie2018}. Leucus was observed less than half a year later by the K2 mission, during the second half of Campaign 11 (in C112 or C11b). Unfortunately, the allocated pixels only provided approximately 300 hr of coverage, i.e., less than a full rotation. Nevertheless, we were able to confirm the 445.73 hr rotation period of Leucus, even from the partial light curve. Preliminary results from stellar occultations\footnote{\url{https://www.lpi.usra.edu/sbag/meetings/jan2019},\\ \url{https://seti.org/leucus-asteroid-efficient-and-accurate\\-citizen-science-unistellar-evscope}} indicate that Leucus has a non-ellipsoidal shape but it is not a binary (although the observations do not rule out a distant companion). The light curve shape and the long period raises the possibility that this asteroid may be a dissociated binary proposed by \citet{nesvorny2020}.

We compared the K2 data to the photometry published by \citet{Buie2018} in Fig.~\ref{fig:Leucus}. We found virtually no difference in the light curve amplitude, but this was not unexpected since a time span of about 7 rotations separated the two measurements and the ecliptic longitude of Leucus changed by only about $10\deg$ during that time. We did find a small but significant phase shift of $\approx0.02$ between the two phase curves. This difference is most likely caused by the different viewing geometries between the Earth and \textit{Kepler}.

\begin{figure}
\includegraphics[width=0.95\columnwidth]{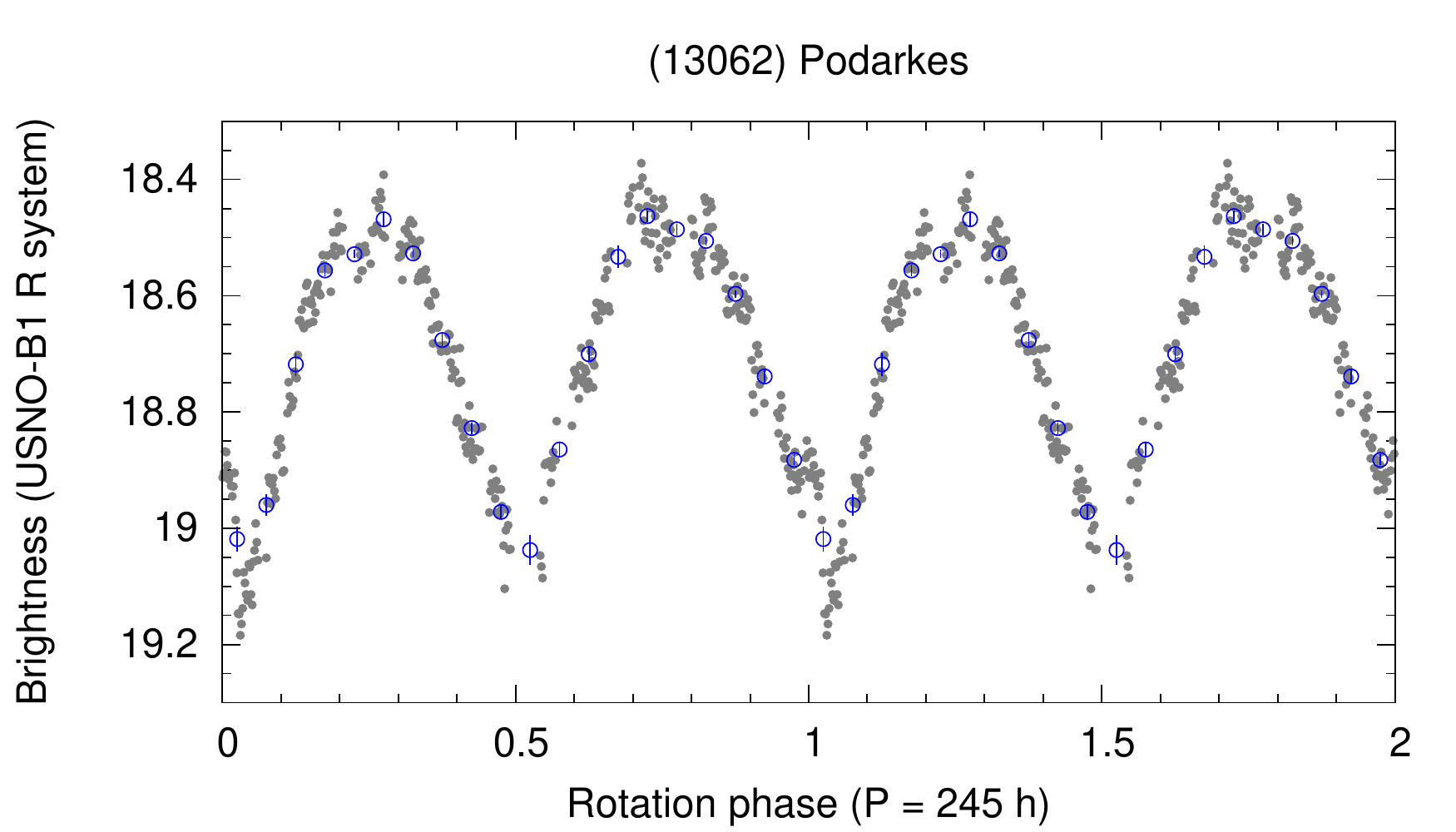}
\caption{Phased-folded light curve of (13062) Podarkes, from Campaign 11. }
\label{fig:podarkes}
\end{figure}

 \subsubsection{(13062) Podarkes}
(13062) Podarkes is the principal body of the proposed Podarkes family that belongs to the larger Menelaus clan \citep{Roig2008}. However, the existence of this family, and other proposed, smaller ones, is not universally accepted, and they do not appear in the general study of asteroid families by \citet{nesvorny2015}. Not much has been known about Podarkes itself before, and no light curve or rotation data can be found in the LCDB. We determined its rotational period to be 245 hr, which means it is another slow rotator (Fig.~\ref{fig:podarkes}). The light curve has a fairly large amplitude of 0.645 mag, with very pronounced, V-shaped minima. 

\begin{figure}
\includegraphics[width=0.95\columnwidth]{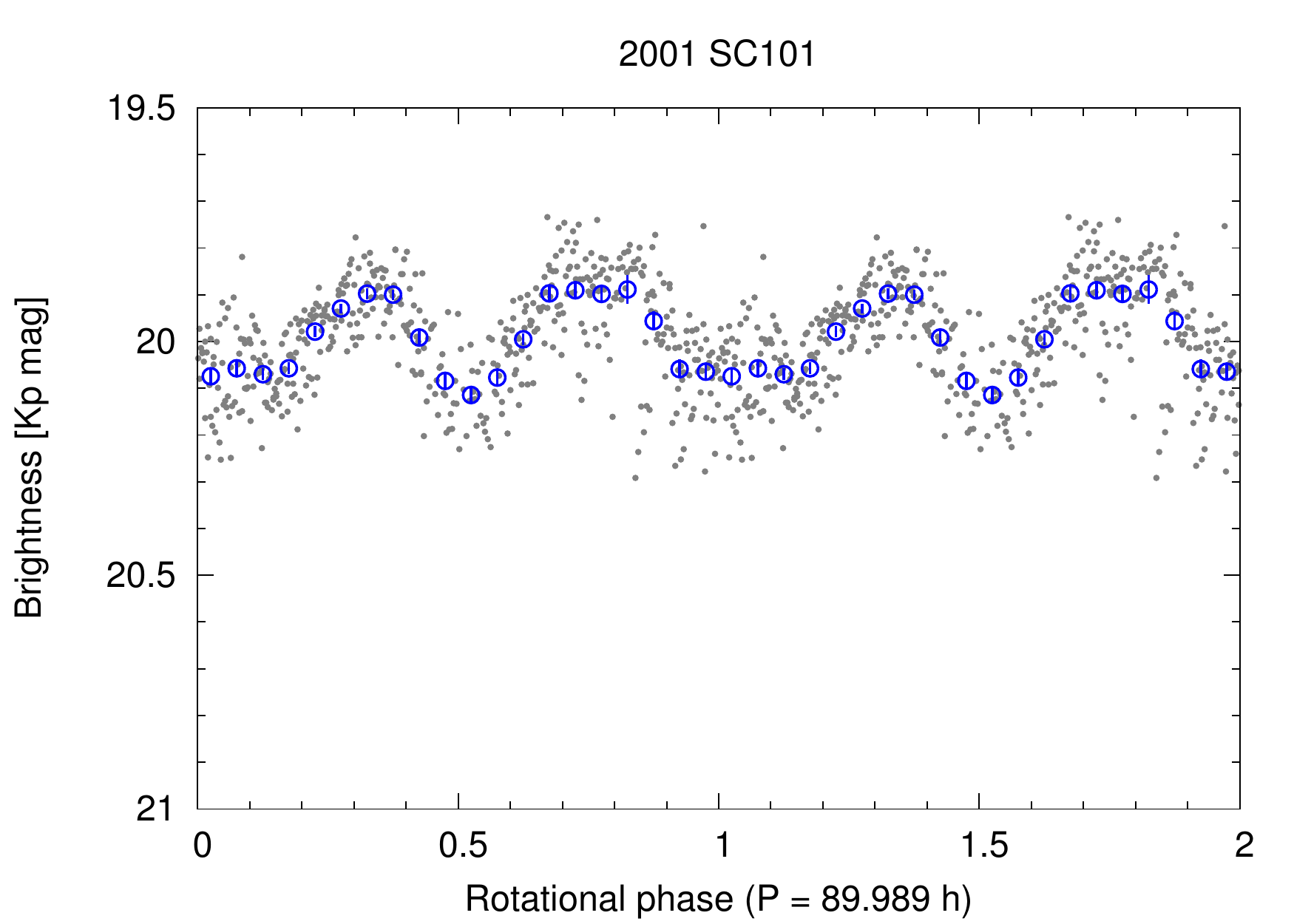}
\includegraphics[width=0.95\columnwidth]{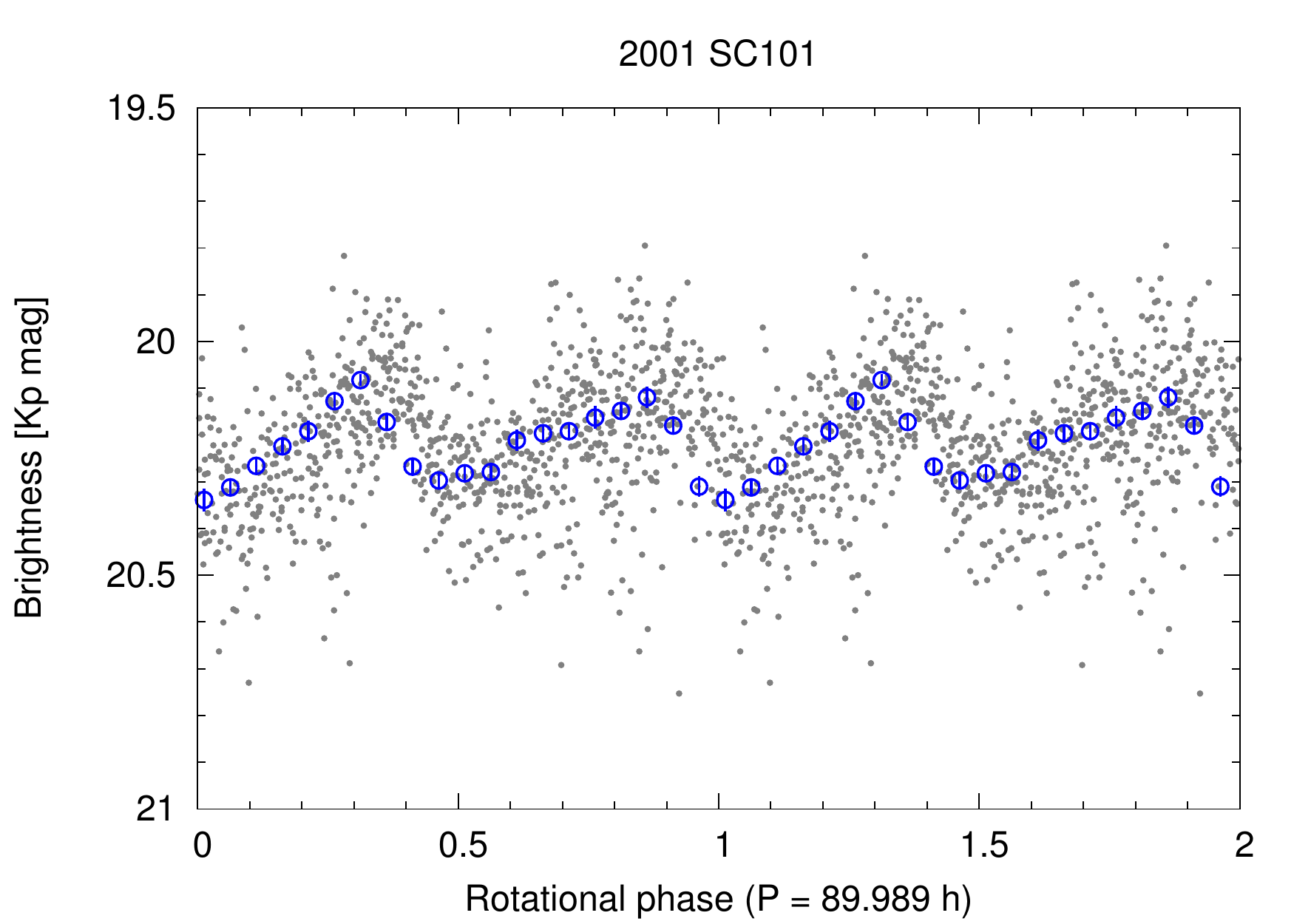}
\caption{Phased-folded light curves of 2001 SC101 as seen by Kepler in C16 (\textit{upper panel}) and C18 (\textit{lower panel}). The rotational period which was used to fold these light curves is the average of the two independent periods obtained from each Campaigns. The two light curves are in the same phase.}
\label{fig:2001SC101}
\end{figure}

\subsubsection{(99306) 2001 SC101}
This is the only Trojan asteroid that has been observed in two different Campaigns during the mission. The fields-of-view of Campaigns 16 and 18 largely overlapped but C16 was observed in forward-facing mode and then C18 in the normal, backward-facing mode. During the 105-day gap between the measurements of (99306) 2001 SC101, the telescope traveled to the other side of the Sun, changing the phase angle it saw the asteroid at from $-10^{\circ}$ to $+10^{\circ}$. The solar elongation of the object also changed by $10.2^{\circ}$. The shape of the light curve changed quite noticeably between the two observations. The amplitudes also differ slightly, but close to the level of uncertainties. Unfortunately, these two epochs are too close in elongation to meaningfully constrain the shape and rotational axis of (99306), and no other time-resolved photometry has been published.

\section{Summary}
\label{sect:summary}
In this paper, we determined photometric properties of 45 Jovian trojans observed in the K2 mission, and present phase-folded light curves for 44 targets. We increased our sample to 101 asteroids with previous K2 Trojan measurements from C6 \citep{Szabo2017}, and derived their amplitude- and frequency distributions. These statistics can provide information on the binary fraction and the composition, which then can be compared to predictions of several formation models. We compared the rotation frequency distribution of the Trojan sample to that of other types of asteroids, like main belt asteroids or Hildas, which we previously determined based on \textit{Kepler} \citep{Szabo2017, Szabo2020ApJS} and TESS data \citep{Pal2020ApJS}. We arrived to the following conclusions:
\begin{itemize}
    \item While 38\% of the periods of K2 Trojans are shorter than 10 hr, 25\% of the sample are slow rotators (P$\geq$30 hr), 12\% are very slow rotators with periods between 100--600 hr.
    \item Compared to other types of asteroids observed by \textit{Kepler} and TESS, we find slow rotators among Hildas (39\%) and main belt asteroids (35\%) too. However, there is no Trojan with rotational period less than 5 hr.
    \item There is a possible dichotomy among Trojans and Hildas in the rotational periods. In the case of Trojans, the short-period group can be fitted with a Maxwellian distribution, which describes a collisionally evolved sample, while the long-period group clearly falls outside the distribution. 
    \item The lack of fast rotators can be interpreted as Trojans generally having low densities, suggesting icy and/or porous composition. Rotation rate statistics indicate that collisional evolution itself would be able to spin Trojans up to $\sim$3--5~hr rotation periods.  We find that deceleration via the YORP effect is very inefficient at this distances, therefore some other process, such as rotational breakup is needed to remove these asteroids from the population. Similarly, the spin barrier based on the photometric properties indicates a critical density upper limit for the Trojans at 0.5 g cm$^{-3}$, which is similar to the TNO population or the nucleus of comets, supporting the scenario of outer Solar System origin.
    \item Based on period (P$\geq$30 hr) and amplitude limits (A$\geq$0.9 mag), we estimate a $\sim$25\% rate of binaries among our sample, consistent with previous measurements \citep{Szabo2017}.
    \item Assuming equator-on viewing geometry and maximum light curve amplitude, a strengthless body model indicates that densities do not exceed 1.5 g cm$^{-3}$, in agreement with the low densities from the spin barrier. We also characterised the possibly binary nature of the K2 Trojan sample, assuming equally sized, spherical, equal-mass and uniform density (1 g cm$^{-3}$) components. We find 32 systems for which both the separation, $a_{bin}/D_0$\,$\geq$\,2 and the amplitude, $A_{max}$\,$\geq$\,0.25\,mag are large enough to classify them as potential detached binary systems. In the period versus diameter diagram (Fig.~\ref{fig:binarity}c), all of K2 Trojans identified as binary candidates are in a region, where double synchronous systems can be found in the main belt.
    \item (11351) Leucus was chosen as one of the flyby targets of the Lucy mission. We confirm the 445.73 hr rotation period, which was based on extensive ground-based observations by \cite{Buie2018}.
    \item We found that (13062) Podarkes -- which is the principal body of the proposed but not universally accepted Podarkes family \citep{Roig2008} -- also has a very slow rotational period (245 hr).
\end{itemize}

Our results once again show that the advantages of continuous and precise photometry collected by space telescopes extend to Solar System objects and planetary science too. With the TESS primary mission already finished and the first extension under way, huge amounts of new photometric data can be expected: however, the small aperture of TESS will limit its ability to study fainter objects (unless temporal resolution is not required, see \citealt{holman-2019RNAAS}). This puts most of the Trojans out of its reach, and the results obtained by the K2 mission even more valuable.

\acknowledgements 
We wish to thank the anonymous referees for their comments that helped to improve this paper. Our results includes data collected by the K2 mission. Funding for the K2 mission is provided by the NASA Science Mission directorate. Cs.E.K.\ was supported by the \'UNKP-19-2 New National Excellence Program of the Ministry of Human Capacities. L.M.\ was supported by the Pr\'emium Postdoctoral Research Program of the Hungarian Academy of Sciences. The research leading to these results has been supported by the LP2018--7/2020 grant of the Hungarian Academy of Sciences and by the K--125015 and  GINOP--2.3.2--15--2016--00003 grant of the Research, Development and Innovation Office (NKFIH), Hungary. This research has made use of NASA's Astrophysics Data System Bibliographic Services.

\facilities{K2 \citep{Howell2014}, JPL HORIZONS \citep{horizons-1996}}
\software{FITSH \citep{pal2012}, numpy \citep{numpy}, matplotlib \citep{matplotlib}, FAMIAS \citep{Zima2008}, gnuplot}

\appendix
%\label{sec:appendix}

\section*{YORP and collisional timescales \label{sect:timescales}}

The timescale to significantly change the rotation state of the asteroid by the YORP effect can be estimated, following \citet{VC2002} and \citet{CV2004}, as:
\begin{equation}
    \tau_{YORP} = \left| {{df}\over{dt}} \right| f_0 = 180 Myr 
    \bigg({{r_h}\over{2.26\,au}}\bigg)^2 \bigg( {{D}\over{6.5\,km}} \bigg)^2 \bigg({{\rho}\over{2.5\,g\,cm^{-3}}} \bigg)
\end{equation}

To estimate the YORP timescale for Jovian Trojans we used r$_h$ = 5.2\,au, D = the values from the actual size estimates, and a bulk density of  $\rho$\,=\,1\,g\,cm$^{-3}$. There is a considerable uncertainty in the density of the Jovian Trojans, but there are indications that their densities is lower than that of main belt asteroids \citep[e.g.][]{Marchis2006}. 

\citet{Farinella1998} estimated the timescale to change a rotation of an asteroid by collisions, $\tau_{coll}$, in the main belt. Using this scheme, we estimated the $\tau_{coll}$ for the Jovian Trojans considering the different number density of objects in the main belt and in the Jovian Trojan swarms. \citet{Nakamura2008} obtained for the L$_4$ swarm that the number of Jovian Trojans larger than 2\,km is N(D$>$2\,km)$_{L4}$\,$\approx$\,6.3$\cdot$10$^4$, while this is N(D$>$2\,km)$_{MB}$\,$\approx$\,5$\cdot$10$^5$ in the main belt \citep{Bottke2015}. Considering the larger space the main belt asteroids occupy this leads to surface densities of $\sigma$(D$>$2\,km)$_{L4}$\,$\approx$\,6.8$\cdot$10$^{-12}$\,km$^{-2}$ and 
$\sigma$(D$>$2\,km)$_{MB}$\,$\approx$\,5.6$\cdot$10$^{-13}$\,km$^{-2}$. We rescaled the \citet{Farinella1998} $\tau_{coll}$ estimate by using the ratio of these surface density values, assuming the same velocity dispersion ($\Delta$v\,$\approx$\,5\,km\,s$^{-1}$) in the two populations and that the targets and impactors have the same densities. With these assumptions the collisional timescale is:
\begin{equation}
\tau_{coll} = 28\,Myr \cdot \bigg( {{R}\over{10\,km}} \bigg)^{{1}\over{2}} 
\end{equation}

The $\tau_{YORP}$ and $\tau_{coll}$ timescales of our Jovian Trojan sample are very different: $\tau_{coll}$\,=\,20--66\,Myr, while $\tau_{YORP}$\,=\,1--100\,Gyr, i.e. the YORP timescales are 2-3 orders of magnitudes longer. This suggests that YORP does not play a significant role in the spin rate evolution of Jovian Trojans in our sample (D\,$>$\,10\,km), and it cannot slow down asteroids to very slow rotation before their rotational state would be overwritten by collisions.

\section*{Summary tables and plots}
\begin{figure*}
    \centering
    \includegraphics[width=1\textwidth]{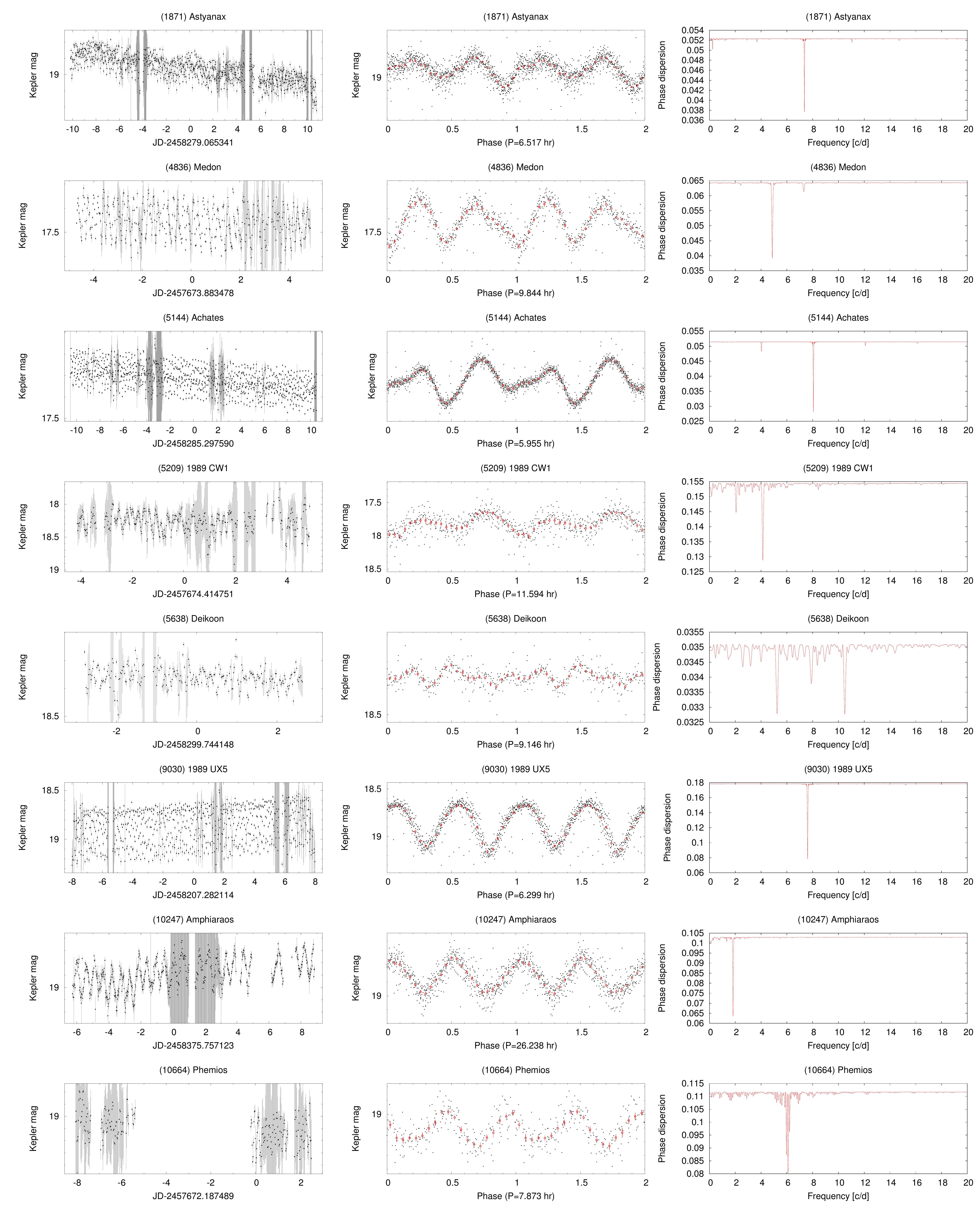}
    \caption{Jovian trojan asteroids observed by the K2 mission between Campaign 11-19. Left: Raw light curves with errors. Middle: Rectified, phase-folded and binned phase curves. Right: Residual dispersion frequency spectra. To fold the raw light curves, we used the half of the main frequency (double-peak solution) in all cases.}
    \label{fig:allplot1}
\end{figure*}

\begin{figure*}
    \centering
    \includegraphics[width=1\textwidth]{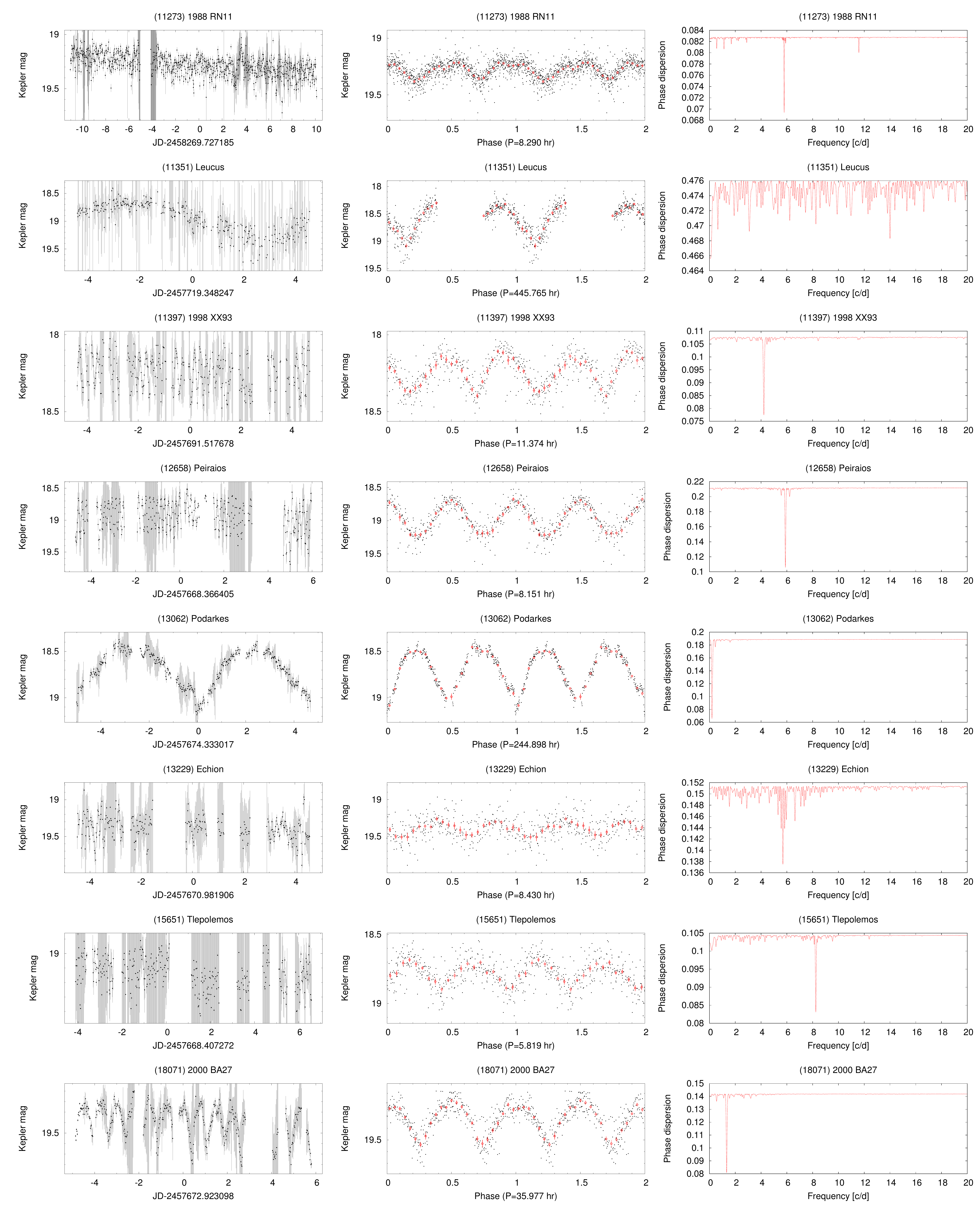}
    \caption{Jovian trojan asteroids observed by the K2 mission between Campaign 11-19, continued. Columns are same as in Fig.~\ref{fig:allplot1}.}
    \label{fig:allplot2}
\end{figure*}

\begin{figure*}
    \centering
    \includegraphics[width=1\textwidth]{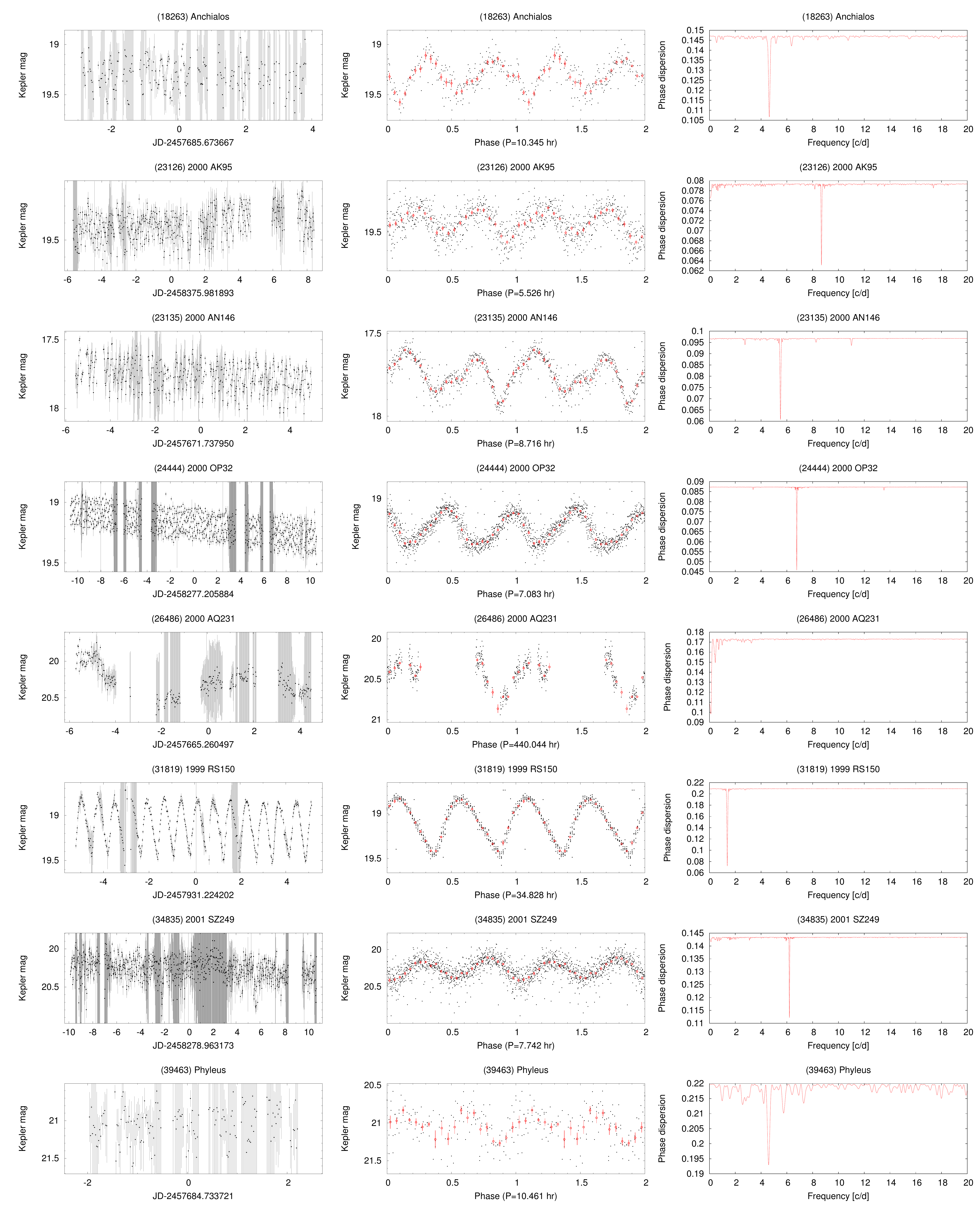}
    \caption{Jovian trojan asteroids observed by the K2 mission between Campaign 11-19, continued. Columns are same as in Fig.~\ref{fig:allplot1}.}
    \label{fig:allplot3}
\end{figure*}

\begin{figure*}
    \centering
    \includegraphics[width=1\textwidth]{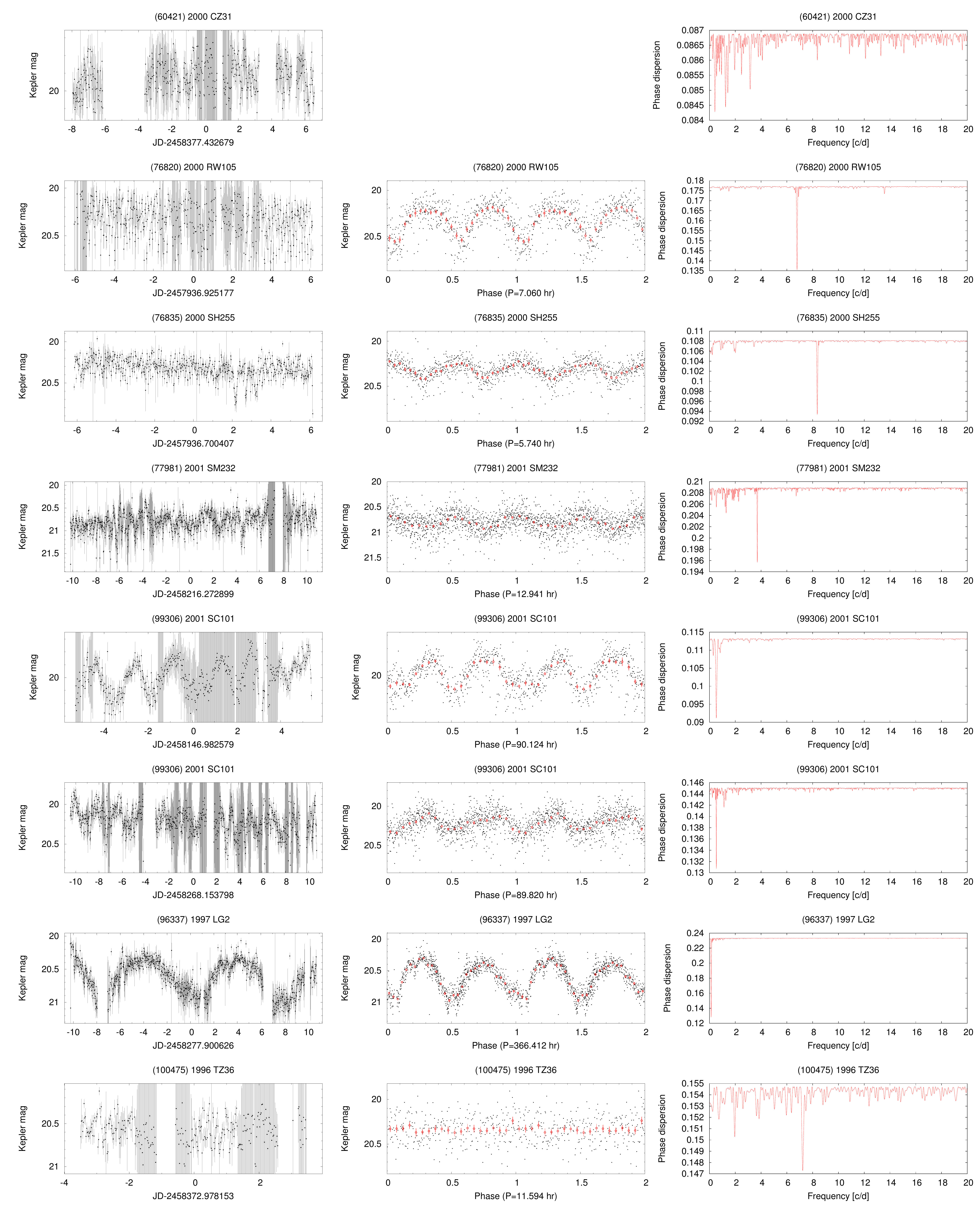}
    \caption{Jovian trojan asteroids observed by the K2 mission between Campaign 11-19, continued. Columns are same as in Fig.~\ref{fig:allplot1}.}
    \label{fig:allplot4}
\end{figure*}

\begin{figure*}
    \centering
    \includegraphics[width=1\textwidth]{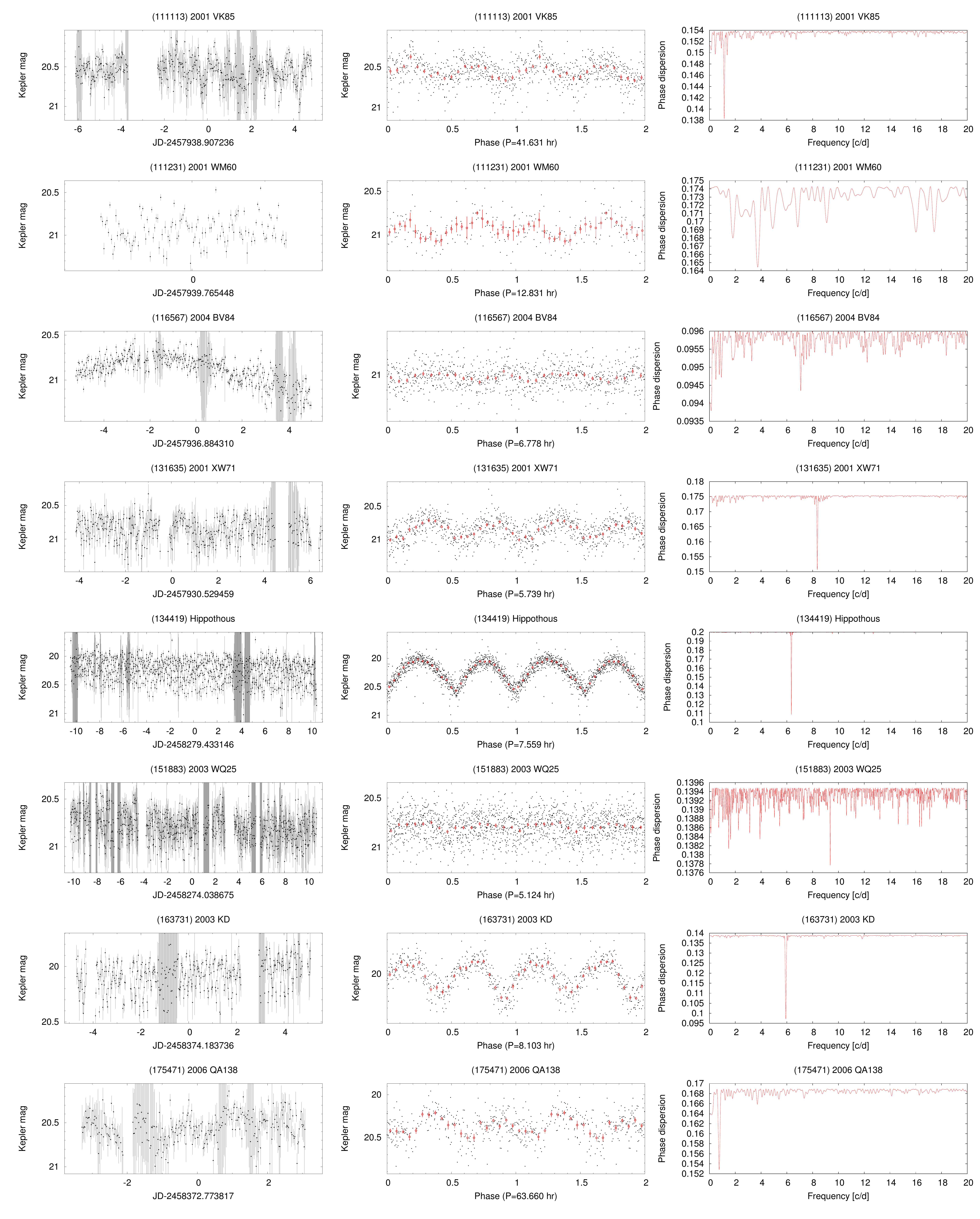}
    \caption{Jovian trojan asteroids observed by the K2 mission between Campaign 11-19, continued. Columns are same as in Fig.~\ref{fig:allplot1}.}
    \label{fig:allplot5}
\end{figure*}

\begin{figure*}
    \centering
    \includegraphics[width=1\textwidth]{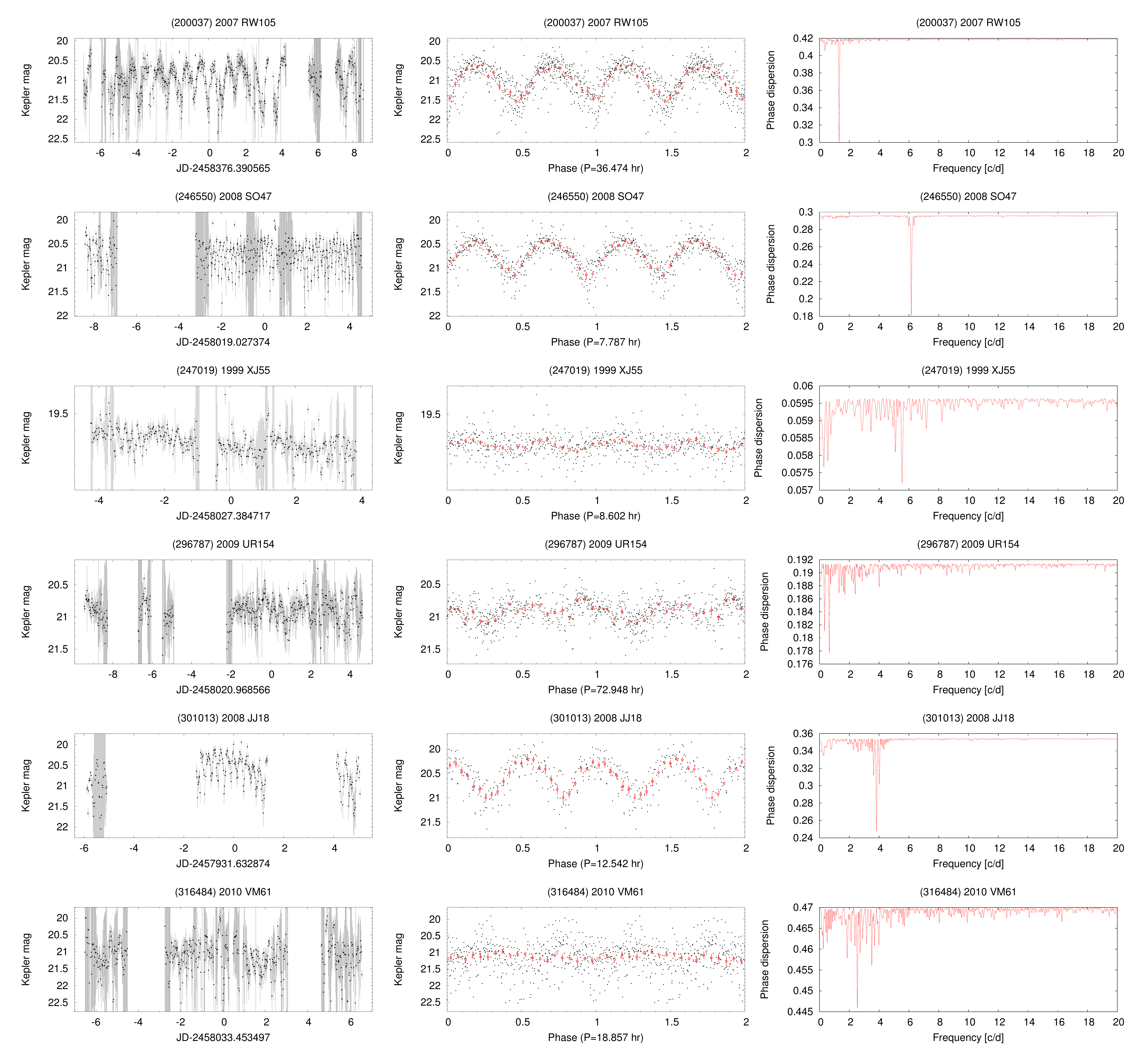}
    \caption{Jovian trojan asteroids observed by the K2 mission between Campaign 11-19, continued. Columns are same as in Fig.~\ref{fig:allplot1}.}
    \label{fig:allplot6}
\end{figure*}

The appendix contains the summary tables and plots. Table \ref{tab:obs} lists the start and end dates of each light curve, the number of data points we used in our analysis, and the duty cycle values. The latter is the ratio of the number of data points we use over the number of data points that would fill the length of the observation. Dates are light receival times at the spacecraft. Table \ref{tab:data} summarizes the results of our analysis. We provide rotation periods and amplitudes plus associated uncertainties, as well as the node designations and campaign numbers for each target. For (99306) 3001 SC101 we list the values obtained from the Campaign 16 and 18 data separately. Finally, in Figures \ref{fig:allplot1}--\ref{fig:allplot6} we display the light curves, phase curves and residual dispersion spectra of each asteroid. 

\startlongtable
\begin{deluxetable*}{lccccc}
\tablecaption{Observation data and duty cycles. Dates are JD--2450000 (d). \label{tab:obs}}
\tablehead{\colhead{Name}  &  \colhead{Start date}  &  \colhead{End date}  &  \colhead{Length (d)}  &  \colhead{No.~points}  &  \colhead{Duty cycle}  }
\colnumbers
\startdata
(1871) Astyanax  &  8268.869  &  8289.936  &  21.07  &  961  &  0.929 \\
(4836) Medon  &  7669.184  &  7678.788  &  9.60  &  453  &  0.961 \\
(5144) Achates  &  8274.754  &  8295.841  &  21.09  &  1022  &  0.987 \\
(5209) 1989 CW1  &  7670.267  &  7679.339  &  9.07  &  364  &  0.817 \\
(5638) Deikoon  &  8296.986  &  8302.360  &  5.37  &  251  &  0.951 \\
(9030) 1989 UX5  &  8199.272  &  8215.333  &  16.06  &  741  &  0.940 \\
(10247) Amphiaraos  &  8369.484  &  8384.605  &  15.12  &  607  &  0.818 \\
(10664) Phemios  &  7664.116  &  7674.701  &  10.58  &  218  &  0.420 \\
(11273) 1988 RN11  &  8258.693  &  8279.842  &  21.15  &  977  &  0.941 \\
(11351) Leucus  &  7714.935  &  7723.925  &  8.99  &  315  &  0.714 \\
(11397) 1998 XX93  &  7687.124  &  7696.258  &  9.13  &  333  &  0.743 \\
(12658) Peiraios  &  7663.646  &  7674.333  &  10.69  &  361  &  0.688 \\
(13062) Podarkes  &  7669.327  &  7679.012  &  9.69  &  374  &  0.787 \\
(13229) Echion  &  7666.487  &  7675.559  &  9.07  &  284  &  0.638 \\
(15651) Tlepolemos  &  7664.300  &  7674.987  &  10.69  &  335  &  0.639 \\
(18071) 2000 BA27  &  7668.060  &  7678.726  &  10.67  &  392  &  0.749 \\
(18263) Anchialos  &  7682.772  &  7689.495  &  6.72  &  231  &  0.700 \\
(23126) 2000 AK95  &  8370.301  &  8384.380  &  14.08  &  523  &  0.757 \\
(23135) 2000 AN146  &  7666.180  &  7676.744  &  10.56  &  468  &  0.902 \\
(24444) 2000 OP32  &  8266.560  &  8287.852  &  21.29  &  903  &  0.864 \\
(26486) 2000 AQ231  &  7659.539  &  7669.776  &  10.24  &  281  &  0.559 \\
(31819) 1999 RS150  &  7926.014  &  7936.353  &  10.34  &  484  &  0.954 \\
(34835) 2001 SZ249  &  8269.094  &  8289.691  &  20.60  &  919  &  0.909 \\
(39463) Phyleus  &  7682.772  &  7686.941  &  4.17  &  151  &  0.738 \\
(60421) 2000 CZ31  &  8369.484  &  8383.992  &  14.51  &  474  &  0.666 \\
(76820) 2000 RW105  &  7930.877  &  7943.076  &  12.20  &  534  &  0.892 \\
(76835) 2000 SH255  &  7930.550  &  7942.851  &  12.30  &  594  &  0.984 \\
(77891) 2001 SM232  &  8206.077  &  8227.184  &  21.11  &  988  &  0.953 \\
(96337) 1997 LG2  &  8267.643  &  8288.669  &  21.03  &  896  &  0.868 \\
(99306) 2001 SC101 (C16) & 8141.690  &  8152.377  &  10.69  &  501  &  0.955 \\
(99306) 2001 SC101 (C18) & 8257.692  &  8278.841  &  21.15  &  895  &  0.862 \\
(100475) 1996 TZ36  &  8369.484  &  8376.431  &  6.95  &  277  &  0.812 \\
(111113) 2001 VK85  &  7932.777  &  7943.750  &  10.97  &  463  &  0.860 \\
(111231) 2001 WM60  &  7938.519  &  7941.053  &  2.53  &  122  &  0.981 \\
(116567) 2004 BV84  &  7931.674  &  7941.870  &  10.20  &  458  &  0.915 \\
(131635) 2001 XW71  &  7926.381  &  7938.478  &  12.10  &  513  &  0.864 \\
(134419) Hippothous  &  8268.930  &  8289.916  &  20.99  &  1014  &  0.984 \\
(151883) 2003 WQ25  &  8263.740  &  8284.787  &  21.05  &  839  &  0.812 \\
(163731) 2003 KD  &  8369.484  &  8379.292  &  9.81  &  400  &  0.831 \\
(175471) 2006 QA138  &  8369.484  &  8375.839  &  6.35  &  289  &  0.926 \\
(200037) 2007 RW105  &  8369.464  &  8384.973  &  15.51  &  583  &  0.766 \\
(246550) 2008 SO47  &  8010.609  &  8023.666  &  13.06  &  446  &  0.696 \\
(247019) 1999 XJ55  &  8023.135  &  8031.247  &  8.11  &  368  &  0.924 \\
(296787) 2009 UR154  &  8011.508  &  8025.668  &  14.16  &  453  &  0.652 \\
(301013) 2008 JJ18  &  7925.748  &  7936.700  &  10.95  &  225  &  0.418 \\
(316484) 2010 VM61  &  8026.935  &  8040.033  &  13.10  &  437  &  0.680 \\
\enddata
\end{deluxetable*}

\startlongtable
\begin{deluxetable*}{lcccccc}
\tablecaption{Rotation periods and amplitudes. For (99306) 2001 SC101, we list results from each campaigns individually (starred values). \label{tab:data}}
\tablehead{\colhead{Name}  &  \colhead{Period (h)}  &  \colhead{$\Delta$P (h)}  &  \colhead{Amplitude (mag)}  &  \colhead{$\Delta$A (mag)}  &  \colhead{Lagrange Node}  &  \colhead{Campaign}}
\colnumbers
\startdata
(1871)	Astyanax	&	6.517	&	0.001	&	0.140	&	0.005	&	L5	&	C18	\\	
(4836)	Medon	&	9.844	&	0.004	&	0.199	&	0.007	&	L4	&	C11	\\	
(5144)	Achates	&	5.955	&	0.0004	&	0.176	&	0.003	&	L5	&	C18	\\	
(5209)	1989 CW1	&	11.59	&	0.005	&	0.347	&	0.032	&	L4	&	C11	\\	
(5638)	Deikoon	&	9.146	&	0.029	&	0.077	&	0.006	&	L5	&	C18	\\	
(9030)	1989 UX5	&	6.299	&	0.001	&	0.488	&	0.011	&	L5	&	C17	\\	
(10247)	Amphiaraos	&	26.24	&	0.017	&	0.264	&	0.011	&	L4	&	C19	\\	
(10664)	Phemios	&	7.874	&	0.009	&	0.260	&	0.018	&	L4	&	C11	\\	
(11273)	1988 RN11	&	8.290	&	0.002	&	0.160	&	0.008	&	L5	&	C18	\\	
(11351)	Leucus	&	445.73	&	15.8	&	0.77	&	0.23	&	L4	&	C11	\\	
(11397)	1998 XX93	&	11.374	&	0.011	&	0.279	&	0.015	&	L4	&	C11	\\	
(12658)	Peiraios	&	8.151	&	0.003	&	0.562	&	0.033	&	L4	&	C11	\\	
(13062)	Podarkes	&	245	&	1.3	&	0.645	&	0.015	&	L4	&	C11	\\	
(13229)	Echion	&	8.43	&	0.010	&	0.259	&	0.034	&	L4	&	C11	\\	
(15651)	Tlepolemos	&	5.819	&	0.002	&	0.230	&	0.015	&	L4	&	C11	\\	
(18071)	2000 BA27	&	36.0	&	0.054	&	0.368	&	0.022	&	L4	&	C11	\\	
(18263)	Anchialos	&	10.345	&	0.034	&	0.426	&	0.050	&	L4	&	C11	\\	
(23126)	2000 AK95	&	5.526	&	0.001	&	0.186	&	0.012	&	L4	&	C19	\\	
(23135)	2000 AN146	&	8.716	&	0.003	&	0.308	&	0.011	&	L4	&	C11	\\	
(24444)	2000 OP32	&	7.083	&	0.001	&	0.220	&	0.006	&	L5	&	C18	\\	
(26486)	2000 AQ231	&	440	&	5.1	&	0.470	&	0.072	&	L4	&	C11	\\	
(31819)	1999 RS150	&	34.828	&	0.016	&	0.595	&	0.005	&	L5	&	C14	\\	
(34835)	2001 SZ249	&	7.742	&	0.001	&	0.306	&	0.017	&	L5	&	C18	\\	
(39463)	Phyleus	&	10.46	&	0.041	&	0.408	&	0.049	&	L4	&	C11	\\	
(60421) 2000 CZ31  &     ---    &    ---    &  $<0.05$  &   ---    &   L4  &  C19 \\
(76820)	2000 RW105	&	7.060	&	0.003	&	0.388	&	0.021	&	L5	&	C14	\\	
(76835)	2000 SH255	&	5.740	&	0.002	&	0.202	&	0.017	&	L5	&	C14	\\	
(77891)	2001 SM232	&	12.945	&	0.008	&	0.229	&	0.024	&	L5	&	C17	\\	
(96337)	1997 LG2	&	366.412	&	1.6	&	0.646	&	0.018	&	L5	&	C18	\\	
(99306)	2001 SC101*	&	90.056	&	0.537	&	0.226	&	0.024	&	L5	&	C16	\\	
(99306)	2001 SC101*	&	89.787	&	0.317	&	0.255	&	0.017	&	L5	&	C18	\\	
(100475)	1996 TZ36	&	6.636	&	0.013	&	0.231	&	0.041	&	L4	&	C19	\\	
(111113)	2001 VK85	&	41.631	&	0.171	&	0.297	&	0.028	&	L5	&	C14	\\	
(111231)	2001 WM60	&	12.83	&	0.202	&	0.317	&	0.033	&	L5	&	C14	\\	
(116567)	2004 BV84	&	6.778	&	0.013	&	0.065	&	0.015	&	L5	&	C14	\\	
(131635)	2001 XW71	&	5.739	&	0.002	&	0.309	&	0.027	&	L5	&	C14	\\	
(134419)	Hippothous	&	7.559	&	0.002	&	0.539	&	0.016	&	L5	&	C18	\\	
(151883) 2003 WQ25  &  5.1243  &  0.0005  &  0.083  &  0.028  &  L5  &  C18 \\
(163731)	2003 KD	&	8.103	&	0.003	&	0.317	&	0.015	&	L4	&	C19	\\	
(175471)	2006 QA138	&	63.66	&	0.994	&	0.323	&	0.036	&	L4	&	C19	\\	
(200037)	2007 RW105	&	36.47	&	0.046	&	0.964	&	0.060	&	L4	&	C19	\\
(246550)	2008 SO47	&	7.787	&	0.002	&	0.742	&	0.047	&	L4	&	C15	\\	
(247019)	1999 XJ55	&	8.602	&	0.019	&	0.070	&	0.017	&	L4	&	C15	\\	
(296787)	2009 UR154	&	72.948	&	0.521	&	0.397	&	0.025	&	L4	&	C15	\\	
(301013)	2008 JJ18	&	12.542	&	0.011	&	0.890	&	0.074	&	L5	&	C14	\\	
    (316484)	2010 VM61	&	18.857	&	0.045	&	0.650	&	0.112	&	L4	&	C15	\\	
\enddata
\end{deluxetable*}

\end{document}